\documentclass[manuscript,screen,nonacm]{acmart}

\usepackage{amsfonts}
\usepackage{algorithmic}
\usepackage{graphicx}
\usepackage{textcomp}
\usepackage{xcolor}
\usepackage{colortbl}
\usepackage{verbatim}
\usepackage{subcaption}
\usepackage{framed}

\definecolor{lightgray}{gray}{0.95}
\definecolor{darkblue}{rgb}{0,0,0.5}
\definecolor{lightred}{rgb}{1, 0.8, 0.8}
\definecolor{lightyellow}{rgb}{1, 1, 0.8}
\definecolor{lightgreen}{rgb}{0.6, 1, 0.6}

\usepackage[commandnameprefix=always,final]{changes}
\setaddedmarkup{\textcolor{blue}{#1}}
\newcommand{\soutblue}[1]{\bgroup\markoverwith{\textcolor{blue}{\rule[0.5ex]{2pt}{0.8pt}}}\ULon{#1}}
\setdeletedmarkup{\textcolor{black}{\soutblue{#1}}}

\DeclareMathOperator*{\argmin}{ArgMin}

\AtBeginDocument{%
  \providecommand\BibTeX{{%
    \normalfont B\kern-0.5em{\scshape i\kern-0.25em b}\kern-0.8em\TeX}}}

\begin{document}

\title{Beyond Dependencies: The Role of Copy-Based Reuse in Open Source Software Development}

\author{Mahmoud Jahanshahi}
\email{mjahansh@vols.utk.edu}
\author{David Reid}
\email{dreid6@vols.utk.edu}
\author{Audris Mockus}
\email{audris@utk.edu}
\affiliation{
  \department{Department of Electrical Engineering and Computer Science}
  \institution{The University of Tennessee}
  \city{Knoxville}
  \state{TN}
  \country{USA}
}

\begin{abstract}
In Open Source Software, resources of any project are open for reuse by introducing dependencies or copying the resource itself.
In contrast to dependency-based reuse, the infrastructure to systematically support copy-based reuse appears to be entirely missing.
Our aim is to enable future research and tool development to increase efficiency and reduce the risks of copy-based reuse.
We seek a better understanding of such reuse by measuring its prevalence and identifying factors affecting the propensity to reuse.
To identify reused artifacts and trace their origins, our method exploits World of Code infrastructure.
We begin with a set of theory-derived factors related to the propensity to reuse, sample instances of different reuse types, and survey developers to better understand their intentions.  
Our results indicate that copy-based reuse is common, with many developers being aware of it when writing code.
The propensity for a file to be reused varies greatly among languages and between source code and binary files, consistently decreasing over time.
Files introduced by popular projects are more likely to be reused, but at least half of reused resources originate from ``small'' and ``medium'' projects.
Developers had various reasons for reuse but were generally positive about using a package manager.
\end{abstract}




% to do: should be updated

\maketitle

\section{Introduction}
% 1. code reuse in general
Software reuse refers to the practice of developing software systems from existing software rather than creating them from scratch~\cite{krueger1992software}. 
Starting from scratch may demand more time and effort than reusing pre-existing, high-quality code that fits the required task.
Developers, therefore, opportunistically and frequently reuse code~\cite{juergens2009code}.
Programming for clearly defined problems often starts with a search in code repositories, typically followed by careful copying and pasting of the relevant code~\cite{sim1998archetypal}.

% 2. OSS development, its platforms and how they facilitated code reuse
The fundamental principle of Open Source Software (OSS) lies in its ``openness'', which enables anyone to access, inspect, and reuse any artifact of a project. 
This could significantly enhance the efficiency of the software development process. 
Platforms such as GitHub significantly increase reuse opportunities by enabling the community of developers to curate software projects and by promoting and improving the process of opportunistic discovery and reuse of artifacts~\cite{jahanshahi2023dataset}.
A significant portion of OSS is intentionally built to be reused, offering resources or functionality to other software projects~\cite{haefliger2008code}, thus such reuse can be categorized as one of the building blocks of OSS. 
Indeed, developers in the open source community not only seek opportunities to reuse existing high-quality code, but also actively promote their own well-crafted artifacts for others to utilize~\cite{gharehyazie2017some}. 
Being widely reused not only increases the popularity of the software project and its maintainers while providing them with job prospects~\cite{Roberts06}, but also may bring new maintainers as well as corporate support~\cite{jahanshahi2023dataset}.

% type of reuse
Most commonly, code reuse refers to the introduction of explicit dependencies on the functionality provided by ready-made packages, libraries, frameworks, or platforms maintained by other projects (referred to as dependency-based or black-box reuse). 
Such external code is not modified by the developer and, generally, not committed into the project’s repository but relied upon via a package manager. 
Copy-based reuse (or white-box reuse), on the other hand, refers to the case where source code (or other reusable artifacts) is reused by copying the original code and committing the duplicate code into a new repository. 
It may remain the same or be modified by the developer after reuse. 
We specifically focus on copy-based reuse in this study.

% parallel evolution of code, the package manager idea
While it is generally accepted that programs should be modular~\cite{parnas1972criteria}, with internal implementation details not exposed outside the module, copy-based reuse does exactly the opposite.
OSS's copy-based reuse, where any source code file or even a code snippet can be reused in another project, may result in multiple, possibly modified instances of the same source code replicated across various files and repositories. 
These copies may undergo further changes during maintenance, leading to multiple different versions of the originally identical code existing in the latest releases of corresponding projects. 
Unifying such multiplicity of versions in copy-based reuse to refactor it into a single package that all these projects could depend upon may not always be a tractable problem. 

% attribution loss/legal issues
Moreover, as this reuse process continues across various projects, possibly with some modifications, data related to the initial design, authorship, copyright status, and licensing could be lost~\cite{qiu2021empirical}.
This loss could impede future enhancements and bug-fixing efforts. 
It might also diminish the motivation for original authors who seek recognition for their work and lead to legal complications for downstream users.
These issues impact not only those who reuse the code but also the software dependent on at least one package that involves reused code~\cite{8667977}.

% poor quality, security, ... spread
As the landscape of Open Source Software (OSS) expands, tracing the origins of source code, identifying high-quality code suitable for reuse, and deciphering the simultaneous progression of code across numerous projects become increasingly challenging.
This can pose risks, such as the spread of potentially low-quality or vulnerable code~\cite{jahanshahi2023dataset} (e.g, orphan vulnerabilities~\cite{9794064}).

% motivation
Despite the sustained attention and potential substantial benefits and risks associated with reuse, the exact scale, prevalent practices, and possible negative impacts related to OSS-wide reuse have not been thoroughly explored.
This is primarily due to the formidable task of tracking code throughout the entirety of OSS~\cite{jahanshahi2023dataset}.

Gaining a more comprehensive understanding of reuse practices could guide future research towards developing methods or tools that enhance productivity while mitigating the inherent risks associated with reuse. 
Specifically, we aim to quantify several aspects concerning the extent and nature of reuse in OSS, providing information necessary to investigate approaches that support this common activity, making it more efficient and safer.

We use a measurement framework created in~\cite{jahanshahi2023dataset} that tracks all versions of project artifacts, referred to as blobs\footnote{
In alignment with the terminology used in the Git version control system, we use the term ``blob'' to refer to a single version of a file.},
across all repositories.
\chadded{
In this approach, the first time each blob is committed to a repository is identified. 
The (repository, blob) tuples are then sorted based on the commit time of the first appearance of that unique blob in the repository.
The repository with the earliest commit time is identified as the originating repository, and the person who made that commit is recognized as the creator of the blob.
Reuse instances are then identified by pairing the originating repository with any subsequent repositories that commit the same blob.
}

% outline
Our work investigates how much and what kind of reuse happens at the scale of OSS, with findings that could help guide future research and tool development to support this common but potentially risky activity.
First, we show how the existing studies, by ignoring ``small'' and inactive projects, miss almost half of the code reused even by the ``largest'' and most active projects.
There is a necessity for more in-depth study to fully comprehend how these abundant yet unseen ``dark matter'' projects significantly contribute to reuse activity.
Second, we theorize about and investigate empirically the properties of artifacts and originating projects that influence the likelihood of file reuse, addressing a key question that previous work, which has predominantly focused on copy detection techniques, has missed.
To investigate historic reuse trends, we also introduce a time-limited measure of reuse. 
Our findings reveal several surprising patterns showing how copying varies with the programming language, properties of a blob, and originating projects.
These insights could help prioritize and articulate further research and tool development that supports the most common reuse patterns.
Third, we obtain responses from 374 developers about the code they have reused or originated.
Most respondents write code with an explicit expectation that it will be reused. 
Developers reuse code for several reasons and are not concerned with bugs in the reused code, but they are willing to use package managers for reused code if such tools were provided. 
Overall, we find that despite its questionable reputation due to inherent risks, code copying is common, useful, and many developers keep it in mind when writing code.

In summary we ask the following research questions:
\begin{enumerate}
    \item[RQ1] How much copy-based reuse occurs? What factors affect the propensity to reuse?
    \begin{enumerate}
        \item \label{rq1-a} How extensive is copying in the entire OSS landscape?
        \item \label{rq1-b} Is copy-based reuse limited to a particular group of projects?
        \item \label{rq1-c} Do characteristics of the blob affect the probability of reuse?
        \item \label{rq1-d} Do characteristics of the originating project affect the probability of reuse?
    \end{enumerate}
    \item[RQ2] \chadded{How do developers perceive and engage with copy-based reuse?}
\end{enumerate}

\section{Background} \label{Reuse}
\chadded{
This section is structured to provide a comprehensive understanding of the context and foundation for our research. 
It begins with an exploration of the types of reuse in software supply chains.
Following this, we delve into the associated risks, discussing potential vulnerabilities, legal issues, and other challenges that can arise from software reuse. 
The third subsection introduces the social contagion theory (SCT), which serves as the theoretical framework for our analysis. 
This theory helps explain the diffusion and adoption of reuse practices within the open source software development community. 
}

\subsection{Reuse in Software Supply Chains}
\chadded{
A software supply chain comprises various components, libraries, tools, and processes used to develop, build, and publish software artifacts. 
It covers all stages from initial development to final deployment, including proprietary and open source code, configurations, binaries, plugins, container dependencies, and the infrastructure required to integrate these elements. 
The software supply chain ensures that the right components are delivered to the right places, at the right times, and in the correct quantities to create functioning software products.
Integrating software reuse within the supply chain enhances efficiency, reduces costs, and mitigates the risks associated with developing new software from scratch.
}

In the context of open source software, reuse in supply chains can be categorized based on how the open source components are integrated and utilized within software projects~\cite{fse19,isectut22,ase23}.

\chadded{
\textit{Dependency-Based Reuse.} This type involves using open source libraries and packages as dependencies in a project. 
These dependencies are typically managed through package managers such as NPM for JavaScript, pip for Python, or Maven for Java. 
The reliance on these dependencies can introduce vulnerabilities and risks if not properly managed~\cite{9724801}.
A web application using the React library, which in turn depends on numerous other libraries is an example of reuse in this kind of supply chain.
}

\chadded{
\textit{Copy-Based Reuse.} It is the type of reuse investigated in this work.
In copy-based reuse, code from open source projects is copied directly into a project. 
For example, a developer might copy a utility function from an open source repository and integrate it into their own project.
While this approach is quick, it can lead to challenges in maintaining and updating the copied code. 
It is essential to track and manage these copies to ensure they are secure and up-to-date~\cite{10179304}.
}

\chadded{
\textit{Knowledge-Based Reuse.} This type involves using knowledge and practices derived from open source projects without directly copying code or using dependencies. 
It includes the adoption of development methodologies, architectural patterns, and best practices from open source communities.
For example, implementing a microservices architecture inspired by successful open source projects.
While not explicitly detailed by many researchers, the concept of knowledge-based supply chains is inferred from broader discussions of open source influence on software development practices~\cite{zhao2021evaluation} 
}

\subsection{Associated Risks}
% % 1. why is it important?
While reuse can potentially reduce development costs, it is not always beneficial. 
It could introduce certain risks that might eventually escalate the overall costs of a project. 
These risks include, but are not limited to, security vulnerabilities, compliance, and the spread of bugs or low-quality code~\cite{jahanshahi2023dataset,german2009code}. 

% 2. security
\subsubsection{Security}
The relationship between security and reuse can possess a dual-nature: 
a system can become more secure by leveraging mature dependencies, but it can also become more vulnerable by creating a larger attack surface through exploitable dependencies~\cite{gkortzis2021software}. 

In the context of copy-based reuse, extensive code copying can lead to the widespread dissemination of potentially vulnerable code. 
These artifacts may reside not only in inactive projects (that are still publicly available for others to reuse and potentially spread the vulnerability further), but also in highly popular and active projects~\cite{9794064}.

\chadded{
Understanding the copy-based supply chain helps in identifying potential security risks and implementing appropriate safeguards~\cite{10.1145/3560835.3564556}.
Therefore, detecting reused code aids in identifying and consistently patching these vulnerabilities across all affected systems~\cite{10179304}.
}

% 3. licensing
\subsubsection{Compliance}
\chadded{
Many open-source licenses come with specific requirements that must be met. 
Unintentional reuse of code that is subject to intellectual property (IP) rights or licensing restrictions can lead to legal complications.
Understanding the supply chain and detecting reused artifacts ensures compliance with licensing agreements and protects against IP infringements~\cite{LIANG2022106,zhao2021evaluation}.
}

As software systems evolve, their licenses evolve as well. 
This evolution can be driven by various factors such as changes in the legal environment, commercial code being licensed as free and open source, or code that has been reused from other open source systems. 
The evolution of licensing can impact how a system or its parts can be subsequently reused~\cite{jahanshahi2023dataset}. 
Therefore, monitoring this evolution is important~\cite{di2010exploratory}. 
However, keeping track of the vast amount of data across the entire OSS landscape is a challenging task, and as a result, many developers fail to adhere to licensing requirements~\cite{an2017stack,german2009license}.

For example, investigating a subset of codes reused in the Stack Overflow environment revealed an extensive number of potential license violations~\cite{an2017stack}. 
Even when all license requirements are known, the challenge of combining software components with different and possibly incompatible licenses to create a software application that complies with all licenses, while potentially having its own, persists and is of great importance~\cite{german2009license}. 
When individual files are reused, licensing information may be lost, and the findings of our study might suggest approaches to identify and remediate such problems.

% 4. quality/bugs
\subsubsection{Quality}
\chadded{
Ensuring that all components of the supply chain meet quality standards is essential for the reliability and performance of the final product~\cite{10.1145/3639476.3639775}.
Copied code that has not been thoroughly vetted and tested can introduce bugs and defects. 
By identifying and evaluating such reused code, organizations can ensure that it meets their quality standards~\cite{fse19}.
}

Code reuse is not only assumed to escalate maintenance costs under specific conditions, but it is also seen as prone to defects. 
This is because inconsistent modifications to duplicated code can result in unpredictable behavior~\cite{juergens2009code}. 
Additionally, failure to consistently modify identifiers (such as variables, functions, types, etc.) throughout the reused code can lead to errors that often bypass compile-time checks and transform into hidden bugs that are extremely challenging to detect~\cite{li2006cp}. 

Apart from the bugs introduced through code reuse, the source code itself could have inherent bugs or be of low quality. 
These issues can propagate similarly to how security vulnerabilities spread.
The patterns of reuse identified in this study could potentially suggest strategies to leverage information gathered from multiple projects with reused code, thereby reducing such risks.

\subsection{Social Contagion Theory}
Reusing code is an instance of technology adoption. 
One of the key questions we want to ask is what may affect the propensity of adopting (copying) a blob.
Social Contagion Theory (SCT)~\cite{Christakis2013} is a widely used theoretical framework for examining dynamic social networks and human behavior in the context of technology adoption~\cite{angst2010social,samadi2016subjective}.
In the field of software engineering, it has been used to explain how developers select software packages~\cite{9091014}.

\chadded{
We are using SCT to theorize about the dynamics of code reuse by conceptualizing it in terms of exposure, infectiousness, and susceptibility. 
SCT helps us frame our research questions by providing a structured way to analyze how code reuse spreads within the open source community.
Specifically, we explore how developers become aware of reusable code, the inherent qualities of the code that make it more likely to be reused, and the characteristics of projects or developers that make them more likely to adopt reusable code.
These dimensions guide the formation of our research questions, enabling us to systematically investigate the factors influencing reuse activity in open source software.
}
The key value of SCT in our case is to help articulate factors affecting copy propensity via three dimensions:

\begin{enumerate}
  \item {\it Exposure}: An intuitive notion that in order to copy the blob, you first have to learn about and find it.
  \item {\it Infectiousness}: The property of the artifact that affects its propensity to be reused.
  \item {\it Susceptibility}: The property of the destination project or developer that reflects how much benefit they would (or believe they would) derive by reusing the artifact.
\end{enumerate}

%We discuss how to operationalize these intuitive concepts below.
First, for a blob (infectious agent) to be reused, a developer needs to become aware of it. 
In other words, it needs to be exposed to the open source community (population). 
Social coding platforms such as GitHub provide various crowd-sourced signals of project popularity.
Developers may consider these characteristics of project popularity or health when choosing what resource to use~\cite{frakes1996software,lozano2002baremo}.
These considerations suggest that developers are more likely to be exposed to code in more popular or active projects.
Therefore, we used project properties as a proxy for the likelihood of awareness. 
This primarily addresses RQ1-b and RQ1-d in our study.

The second concept of SCT, infectiousness, means that a highly virulent infectious agent is more likely to spread.
In our context, this can be measured by the characteristics of the blob itself, corresponding to RQ1-c.
Most of the literature on reuse has primarily focused on this aspect of the reused resource.

The final concept in our theoretical framework is susceptibility, which refers to the vulnerability of the target population to the infectious agent.
In our case, this can be approximated by the characteristics of the target project (or author) that reuses the blob.
For example, the use value, or how much the blob is needed in the project that copies it.
These characteristics are, by definition, highly specific to the target project, making them more challenging to measure.
We aim to shed more light on this aspect in RQ2.

\section{Related Work and Contributions}
While the benefits and risks associated with code reuse seem tangible, the extent and types of reuse across the entirety of OSS remain unclear. 
To prioritize these risks and benefits, and explore methods to minimize or maximize them respectively, we employ the approach introduced in our previous work~\cite{jahanshahi2023dataset}.
This method allows us to track copy-based reuse on a scale commensurate with the vast size of OSS.
The scope of copying activity is not fully encompassed by previous studies based on convenience samples, as we will illustrate in the results section.

We are not aware of any other curation system that operates at the level of a blob or finer granularity, nor is there an easy way to determine the extent of OSS-wide copy-based reuse at that level.
Methods for identifying reuse, such as the one introduced in~\cite{6975664}, are designed to find reuse between specific input projects and do not easily scale to detect reuse across all OSS repositories~\cite{jahanshahi2023dataset}. 
The methods we use to identify and characterize reuse could, therefore, serve as a foundation for tools that expose this difficult-to-obtain yet potentially important phenomenon~\cite{jahanshahi2023dataset}. 
We acknowledge that the actual extent of reuse is most likely much higher than what we find at blob-level granularity.
Nevertheless, we believe the results we present will still be insightful, especially as the lower bound for the extent of copy-based reuse activity in the entirety of OSS.

\chadded{We first differentiate copy-based reuse from related fields and then discuss our contributions.}

\subsection{Related Research Areas}
% contributions
\chadded{
To comprehensively understand copy-based reuse, it is essential to discuss two closely related fields: the clone detection and the clone-and-own practice.
Following discussion will focus on differentiating copy-based reuse from dependency-based reuse, clone detection, and clone-and-own practices, situating these within the broader context of code reuse literature.
}

\subsubsection{Code Reuse Analysis}
\chadded{
It encompasses techniques and practices that aim to maximize the efficiency and reliability of software development by leveraging existing code.
Techniques such as static analysis, dependency analysis, and repository mining help identify reusable components within a codebase~\cite{koschke2007survey}.
Through these methods, code reuse analysis seeks to reduce redundancy and enhance maintainability.
Research indicates that systematic code reuse can significantly reduce development time and costs while improving software quality~\cite{frakes2005software}.
}

\subsubsection{Clone Detection}
\chadded{
It is a technique within code reuse analysis for identifying similar or identical code fragments in a codebase.
This process involves using tools to detect exact or slightly modified duplicates, which can then be refactored into reusable components.
Techniques range from textual and token-based methods to more advanced semantic and abstract syntax tree (AST) analyses~\cite{roy2007survey,svajlenko2014evaluating}.
These methods focus on identifying code clones within constrained contexts, often limited to small code snippets within a few projects~\cite{svajlenko2015evaluating}.
Clone detection helps in managing redundancy and maintaining code quality by highlighting areas where code can be simplified and reused~\cite{roy2007survey}.
The effectiveness of clone detection tools has been validated in various studies, showing significant improvements in software maintainability~\cite{kapser2008cloning}. 
}

\subsubsection{Clone and Own}
\chadded{
It is a practice where existing software components are copied and modified to meet new requirements.
This approach is often utilized in product line engineering and situations where rapid development is critical.
Clone-and-own allows developers to quickly adapt existing solutions but can lead to maintenance challenges due to the proliferation of similar, independently maintained code fragments~\cite{krueger2001easing,rubin2013survey}.
This practice, common in open source development, involves significant modifications and independent maintenance, often leading to divergent development paths~\cite{german2002evolution,blincoe2016understanding}.
}

\chadded{
While clone detection focuses on technical identification of code snippets, the clone-and-own practice highlights the importance of customization and independent management of forked projects.
As the clone-and-own practice involves both technical customization and significant social factors, such as community engagement and governance models., understanding these aspects is crucial for managing forked projects~\cite{german2002evolution,blincoe2016understanding}.
Although clone-and-own supports the purpose of code reuse by facilitating quick adaptation, it often results in code duplication, complicating long-term maintenance.
Research has shown that clone-and-own is prevalent in practice due to its simplicity and effectiveness in the short term~\cite{antoniol2004automatic}.
}

\subsubsection{Copy-based Reuse}
\chadded{
Copy-based reuse, a form of code reuse, involves copying existing code and potentially modifying it for use in new contexts.
This method allows for rapid development but shares the maintenance challenges associated with clone-and-own, as duplicated code must be managed across different parts of the software.
In summary, code reuse analysis encompasses techniques like clone detection to manage redundancy and practices like clone-and-own to adapt existing code for new purposes. 
While clone detection and code reuse analysis share the goal of improving code quality and maintainability by identifying and managing redundancy, clone-and-own focuses on rapid adaptation rather than efficient redundancy management, despite serving a similar purpose in promoting reuse.
Both copy-based reuse and clone detection address code duplication but differ significantly in their methodologies and scopes.
Copy-based reuse research, as exemplified by our work, provides a broader, ecosystem-level perspective, incorporating social aspects and the characteristics of entire projects.
In contrast, clone detection focuses on the technical identification of code snippets within specific contexts, while the clone-and-own practice emphasizes customization and independent maintenance of forked projects. 
}

\subsection{Contributions}
Our contribution in this work has three aspects as follows.

% 1. Entirety (WoC)
\subsubsection{Scale of Analysis}
Our study leverages the World of Code (WoC) infrastructure to analyze reuse at the ecosystem level, encompassing nearly the entire open source software landscape.
\chadded{This comprehensive approach contrasts with previous studies that often focused on specific communities or subsets of programming languages and repositories.}
They either have mostly concentrated on a specific community (e.g. Java language, Android apps, etc.)~\cite{heinemann2011extent,haefliger2008code,mockus2007large,hanna2012juxtapp,7958574,sojer2010code}or only sampled from a single hosting platform (e.g. GitHub)~\cite{gharehyazie2017some, gharehyazie2019cross}.

Even research with more comprehensive programming language coverage~\cite{lopes2017dejavu} analyzed only a subset of programming languages and additionally used convenience sampling methods by excluding less active or ``unimportant'' repositories~\cite{10.1109/ICSE43902.2021.00076,9402500}.
As our results demonstrate, even inactive and ``small'' projects appear to provide many of the artifacts reused in OSS, even by the ``largest'' and most active projects.

\chadded{
Existing literature on code cloning primarily focuses on empirical studies, case studies, and tool evaluations.
Empirical studies typically analyze code clones within specific projects or samples of open-source software repositories.
These datasets are large but not exhaustive of the entire OSS ecosystem.
For example, two studies examine hundreds to thousands of files or repositories, providing valuable but partial insights~\cite{juergens2009code,roy2009comparison}.
Case studies offer in-depth analysis of cloning practices within individual projects or organizations, giving detailed context but limiting the scale to the specific cases under study.
Tool evaluations involve benchmark studies of clone detection tools, evaluating their performance on curated datasets.
While these studies contribute important information about tool effectiveness, they do not cover the entire OSS ecosystem.
}

\chadded{
In contrast, our study stands out due to its unprecedented scale.
Unlike the selective samples analyzed in typical literature, our study encompasses the entire open source software ecosystem, offering a comprehensive view and a exhaustive analysis.
This includes millions of repositories and billions of files.
This scale far surpasses any single empirical or case study in the literature, offering a holistic picture of code reuse practices across the entire OSS landscape.
}

% 2. concentrating on copy-based
\subsubsection{Methodology and Focus}
Copy-based reuse has not been explored as thoroughly as the dependency-based reuse (e.g.,~\cite{cox2019surviving,frakes2001industrial,ossher2010automated}).
\chadded{
For example, studies have shown that dependency-based reuse can lead to more sustainable software architectures by promoting component-based design and reducing redundancy~\cite{mili1995reusing}.
Additionally, it has been demonstrated that by leveraging well-defined interfaces and reusable libraries, dependency-based reuse can significantly improve software maintainability and scalability~\cite{brown1998current}.
Nevertheless, very few, if any, similar analyses exist regarding copy-based reuse.
}
Copy-based reuse is potentially no less important, but is a much less understood form of reuse~\cite{jahanshahi2023dataset}.
Most studies in copy-based reuse domain focus on clone detection tools and techniques~\cite{roy2009comparison,ain2019systematic,jiang2007deckard,hanna2012juxtapp,white2016deep} rather than on the characteristics of entire source code files that possibly make reuse more or less likely.

Furthermore, almost all studies we reviewed focus solely on source code reuse, whereas we track all artifacts, whether they are code or other reusable development resources~\cite{jahanshahi2023dataset}.
By using the World of Code research infrastructure, which encompasses nearly the entire OSS ecosystem, we identified and analyzed copying activity at this scale for the very first time.

\chadded{
In contrast to clone detection, which primarily involves identifying similar code snippets within specific directories or domains~\cite{inoue2021finding,svajlenko2013scaling}, our research addresses the broader context of entire files and diverse artifacts across the OSS ecosystem, providing a more comprehensive understanding of reuse.
Our method bridges the clone detection and clone-and-own approaches by detecting all instances of reuse, whether they are kept without any changes or modified after reuse, thereby encompassing both the technical and managerial aspects of code reuse.
}

\chadded{
In existing clone detection literature, several methods are employed to identify code clones.
These methods include text-based, token-based, tree-based, and graph-based techniques.
Text-based methods detect clones by comparing raw text, which is straightforward but can be less accurate due to variations in formatting.
Token-based methods improve on this by converting code into tokens and detecting similarities at this more abstract level, enhancing accuracy but still being susceptible to variations in code structure.
Tree-based methods parse the code into abstract syntax trees (ASTs) and identify clones by comparing these trees, providing a more structured and semantically meaningful detection.
Graph-based methods further abstract code into control flow or data flow graphs, allowing for the detection of more complex and semantic clones~\cite{roy2009comparison}.
}

\chadded{
The clone and own literature primarily employs these detection methods to understand the broader landscape of code cloning.
For example, a study utilized a combination of these techniques to analyze cloning practices in software projects~\cite{juergens2009code}. 
These methods are effective in identifying different types of clones, such as exact, parameterized, and semantic clones, but they often focus on similarities and patterns rather than exact matches.
}

\chadded{
In contrast, our research employs a method focused on identifying reuse at the blob-level, specifically detecting if the exact versions of code have been copied.
While it misses instances where a single code snippet has been copied, this approach does not rely on abstractions or patterns.
This method involves obtaining hashes for all versions of the entire open source software ecosystem to detect identical code segments, ensuring that every version of code is tracked to its origin.
This exhaustive and detailed approach allows for a comprehensive analysis of copy-based supply chains at the OSS level.
Since software supply chains form a network over the entire OSS, it is not feasible to study them by sampling projects:
representative samples from large graphs are notoriously difficult to obtain (see, e.g.,~\cite{leskovec2006sampling}). 
}

\chadded{ 
In addition to ensuring that the entire file has been copied and committed, our method easily scales to the entire OSS ecosystem as it avoids the need to look for similarities among tens of billions of versions by utilizing hashes.
Traditional clone detection techniques would need to be substantially modified to work at this scale.
We discuss some of the potential approaches in Section~\ref{sec:snippet}.
}

% 3. upstream and downstream projects
\subsubsection{Influencing Factors and Social Aspects}
Our study explores how the characteristics of OSS projects influence the propensity for their artifacts to be reused, examining their social aspects.
Previously, the focus has been primarily on the desired functionality and the code itself~\cite{srinivas2014clustering,geisterfer2006software}, but we also investigate the social aspects of this phenomenon in the open source community.

\chadded{
The literature on clone detection and our research both explore the social aspects of code reuse, but they do so from different perspectives and with varying emphases on social and technical factors.
Existing literature on clone detection primarily focuses on the technical aspects of identifying code clones and understanding their impact on software maintenance and quality.
For instance, studies delve into the reasons for code cloning, such as improving productivity, learning, and avoiding reimplementation of similar functionalities~\cite{juergens2009code,roy2007survey}.
These studies often highlight the technical motivations behind code cloning, such as reusability and rapid prototyping, but they also touch upon social aspects like collaborative development and knowledge sharing within teams.
However, the primary emphasis remains on the technical detection and management of code clones.
}

\chadded{
In contrast, our research takes a broader view by examining how the characteristics of open source software projects influence the propensity for their artifacts to be reused.
This includes a detailed analysis of both social and technical factors.
Our study explores the diverse motivations and implications of reuse in the OSS community, considering aspects such as project size, community engagement, and the collaborative nature of OSS development.
By doing so, we highlight the importance of social dynamics in code reuse, including factors like community contributions, the reputation of projects, and the collaborative environment that fosters code sharing and reuse.
}

\chadded{
By examining these social and technical factors, our study provides a more comprehensive understanding of the motivations behind code reuse in the OSS community.
We draw parallels to other factors influencing copy-based reuse, such as the ease of access to code, the open and collaborative nature of OSS projects, and the role of community support and documentation.
This broader perspective allows us to highlight the diverse and sometimes conflicting motivations for code reuse, ranging from technical efficiency to social recognition and collaborative learning.
}

\section{Methodology}\label{s:method}

We begin by briefly describing the World of Code infrastructure utilized in our study, followed by presenting the methods introduced in our previous work~\cite{jahanshahi2023dataset} to identify instances of copying\footnote{
Online appendix including replication package is available at: https://figshare.com/s/3ad00cceddb820715442}. 
Next, we explain the time complexity of our method and discuss the rationale behind our choice.
In the second and third subsections, we discuss methods used to answer each research question in more detail.

To make the subsequent discussion precise, we first introduce a few definitions. 
The time when each unique {\tt blob} $b$ was first committed to each project $P$ is denoted as $t_b(P)$.
The first repository $P_o(b) = \argmin_Pt_b(P)$ is referred to as the originating repository for $b$ (and the first author as the creator). 
Then project pairs consisting of a project with the originating commit and the destination project with one of the subsequent commits producing the same blob $(P_o(b),P_d(b))$ are identified as reuse instances.
The reuse propensity (the likelihood that a blob will be copied to at least one other project) is then modeled based on the type of the file represented by the blob and the activity and popularity characteristics of the originating projects. 

\subsection{Identification of Reused Blobs} \label{s:copied} 

\subsubsection{World of Code Infrastructure}
Finding duplicate pieces of code and tracking all revisions of that code across all open source projects is a data- and computation-intensive task due to the vast number of OSS projects hosted on numerous platforms~\cite{jahanshahi2023dataset}. 
Previous studies on reuse have consequently often focused on a relatively small subset of open source software, potentially missing the full extent of reuse that could only be obtained with a nearly complete collection~\cite{jahanshahi2023dataset}. 
World of Code (WoC)~\cite{ma2019world,ma2021world} infrastructure aims to address these challenges by regularly discovering, retrieving, indexing, and cross-referencing information from new and updated version control repositories that are publicly available. 

WoC operationalizes copy-based reuse by mapping blobs, which are versions of the source code, to all commits and projects where they have been created.
This means that copy-based reuse is detected only if an entire file is duplicated without any alterations~\cite{jahanshahi2023dataset}. 
If the reuser commits the reused blob before making any modifications, this method will find it;
however, if they commit only after making alterations to the original file, it will not be identified.
Given this, our study focuses solely on whole-file copying activity.
Consequently, different versions of what was originally the same file will be treated as distinct entities since they are different blobs. 

\subsubsection{Project Deforking}
\chadded{
To understand reuse across the entirety of open source software, it is important to identify distinct software projects.
Git commits are based on a Merkle Tree structure, uniquely identifying modified blobs, and therefore, shared commits between repositories typically indicate forked repositories. 
As a distributed version control system (VCS), Git facilitates cloning (via git clone or the GitHub fork button), resulting in numerous repositories that serve as distributed copies of the same project. 
While this feature enables distributed collaboration, it also leads to many clones of the original repository~\cite{mockus2020complete}.}

To differentiate copy-based reuse from forking, we use project deforking map $p2P$ provided in WoC~\cite{mockus2020complete}. 
\chadded{Using community detection algorithms, this map provides a clearer picture of distinct projects by linking forked repositories $p$ to a single deforked project $P$ based on shared commits.}

\chadded{
An advantage of this map over using the fork data from platforms like GitHub is that WoC's p2P map is based on shared commits, providing higher recall by not missing forks that did not occur through GitHub's forking option but rather through cloning the repository.
Additionally, forks and clones hosted on different platforms cannot be traced easily, but the WoC map is platform-independent and does not have this constraint.
Moreover, some forks may diverge significantly from the original repository but are still considered forks by hosting platforms.
WoC's deforking algorithms use community detection via shared commits.
If forks diverge substantially via maintenance after forking, the community detection algorithm would recognize them as distinct projects, which reduces false positives and increases precision.
}

Whenever we mention ``project'' in our paper, we are actually referring to a ``deforked project'' as defined here.
This ensures that our discussions about reuse are based on unique instances of software development projects rather than duplicated efforts through forks.

\subsubsection{Dataset Creation}
To understand the identification of reused blobs, it is important to explain how the dataset we used~\cite{jahanshahi2023dataset} was created. 
Despite the key relationships WoC offers, several critical obstacles had to be resolved. 
The initial step was to pinpoint the first instance, denoted as $t_b(P)$, when each of the approximately 16 billion blobs appeared in each of the almost 108 million projects.
To this goal, first the c2fbb map\footnote{
See https://github.com/woc-hack/tutorial for more information about WoC map naming convention}
(which is the result of diff on a commit: commit file, blob, old blob and lists all blobs created by each commit) was joined with the c2dat map (full commit data) to obtain the date and time of each commit.
The result was then joined with the c2P map (commit to project) to identify all projects containing that commit. 

The result is a new c2btP map (commit to blob,, time, and Project).
To create the timeline for each blob, all that data was sorted by blob, time, and project resulting in b2tP map $(b,t,P)$ where we have only blob, time, and the deforked project that contain our desired timeline $t_b(P)$. 

\begin{figure}
    \centering
    \includegraphics[width=\linewidth]{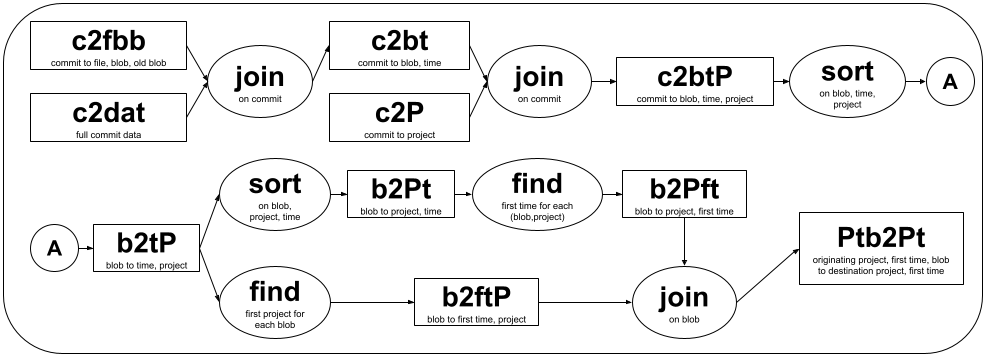}
    \centering
    \caption{Reuse Identification Data Flow Diagram}
    \label{fig:arch}
\end{figure}

Finally, the blob timelines\footnote{
All but the first commit time creating the blob for each project were dropped as a blob is often reused within a repository.}
were used to identify instances of reuse $(t_b(P_o), t_b(P_d))$ or Ptb2Pt map, where the first project is the originating project\footnote{
See section \ref{limit} for the limitations in identifying the originating project.} 
and the second project is the destination project of the reused blob, meaning the blob was created at a later time in this project. 
This resulting Ptb2Pt map contains all instances of blob reuse.
\chadded{The data flow of reuse identification is shown in Figure~\ref{fig:arch}.}

% why stop at blob level?
\subsubsection{Time Complexity Analysis}
\chadded{
To evaluate the complexity and time requirements of our methodology for identifying reuse, we analyze the time complexity of each step and provide a benchmark for execution time on a typical computer setup. 
The overall time complexity is dominated by the sorting operations involved in processing the large maps.
Data preparation and joining involve merging the precalculated maps in WoC, namely the c2fbb, c2P, and c2dat maps. 
Since these maps are already sorted and split into 128 partitions, we can join them with a complexity of $128 \times O(l + m + n)$, where $l$, $m$, and $n$ are the number of rows in the maps respectively. 
We then drop the commit hashes and sort the joined b2tP map based on blob, time, and project, which is the most computationally intensive step, with a complexity of $O(n \log n)$, where $n$ is the total number of rows in the b2tP map. 
Identifying reuse instances, given that the data is already sorted by blob, has a complexity of  $O(n)$, where $n$ is the total number of copy instances.
}

\chadded{
Using a high-performance workstation as a benchmark (8-core processor at 3.5 GHz, 128 GB RAM, 2 TB SSD), we calculate the execution time for each step.
Data preparation and joining, with a linear-time merge, primarily involve reading and writing large files.
With a sequential read/write speed of approximately 500 MB/s for SSDs, joining the maps (total size around 128 billion rows) is expected to take roughly 1-2 hours.
Sorting the created b2tP map, which requires external sorting of about 74 billion rows, necessitates multiple passes over the data.
Based on empirical data, a modern external sorting algorithm with 8 cores can handle around 0.5 billion rows per hour. 
Hence, sorting this map would take approximately 148 hours.
Identifying reuse instances, involving efficient I/O operations, is estimated to take 4-6 hours. 
In total, the entire process is estimated to take approximately 153-156 hours, or about 6.5 days. 
}

\chadded{
Detecting code reuse in finer granularity than blob-level, such as through syntax tree parsing or text similarity techniques, would offer a more comprehensive view of code reuse. 
However, these methods involve several computational challenges and resource constraints, making them impractical for our study.
}

\chadded{
Parsing the abstract syntax tree (AST) for each file to detect structural similarities involves several computational steps. 
First, each file must be parsed into its AST representation, which itself is an $O(n)$ operation where $n$ is the total number of unique blobs.
For our dataset of 16 billion blobs, this parsing step alone would be extremely resource-intensive.
Following parsing, comparing each AST to identify potential reuse instances would require pairwise comparisons. 
The pairwise comparison complexity is $O(n^2)$, resulting in an infeasible $O((16 \times 10^9)^2)$ complexity.
}

\chadded{
Text similarity measures on the other hand, such as Levenshtein distance or cosine similarity, involve comparing each blob's contents with every other blob.
These methods typically operate with a complexity of $O(n^2)$ for each pair of files, again resulting in an infeasible $O((16 \times 10^9)^2)$ complexity.
Even with optimizations like locality-sensitive hashing or other approximation techniques, the scale of the data renders this approach impractical.
}

\chadded{
Given the significant computational complexity and resource requirements, detecting code reuse at a finer granularity than blob-level is not feasible for our study.
Instead, we have chosen to focus on blob-level reuse detection, which provides a practical and scalable solution.
While this approach is limited to detecting exact file copies, it ensures that the analysis remains within the bounds of available computational resources and time constraints, thereby enabling a thorough and efficient examination of code reuse in the OSS landscape.
}

\subsection{RQ1: How much copy-based reuse occurs? What factors affect the propensity to reuse?}

\subsubsection{RQ1-a: How extensive is copying in the entire OSS landscape?} 
To investigate how widespread whole-file copying in OSS actually is, we first want to establish a baseline: 
what fraction of blobs were ever reused, and if reused, to how many downstream projects?
Specifically, in RQ1-a, we are showing the number of blobs, originating as well as destination projects (deforked), and copy instances across the entire OSS ecosystem.
These numbers are not estimates but the actual numbers calculated over the complete dataset.

\subsubsection{RQ1-b: Is copy-based reuse limited to a particular group of projects?}
One may argue that the results in RQ1-a are not necessarily important, as only ``small'' projects may reuse code in a copy-based manner. 
To see if this is actually the case,
\chadded{
we randomly sampled 5 million reuse instances from each of the 128 files into which the data was divided, based on the first two bytes of the hash of blobs.
This resulted in a total of 640 million instances for the analysis.
This approach ensured that our sample was distributed across the entire dataset, capturing a diverse range of copy instances.
The sample size of 640 million instances constitutes approximately 2.67\% of the entire dataset.
Although this is a small fraction of the data, it is sufficiently large to ensure the statistical reliability and representativeness of our analysis, as the large absolute size of the sample guarantees its statistical reliability according to the Central Limit Theorem.
}

\chadded{
Before going further, we need to define the qualitative and, more importantly, subjective terms of ``small'' and ``big'' projects with quantitative and justified measures.
Studies have shown that project activity, as measured by commit frequency, is a strong indicator of project health and sustainability~\cite{crowston2005social,koch2002effort}. 
Additionally, the use of stars as a metric is well-supported in the literature, as they represent a form of user endorsement and are correlated with project visibility and perceived quality~\cite{ray2014large}.
We choose these two metrics because both the number of commits and the number of stars are critical indicators of a project's activity and popularity.
Commits reflect the ongoing development and maintenance efforts, which are crucial for the sustainability and evolution of a project.
Stars, on the other hand, reflect the community's interest and endorsement, indicating the project's visibility and influence.
These metrics are widely used in empirical software engineering research to evaluate the health and impact of open source projects~\cite{jiang2007deckard,borges2016predicting}.
}

\chadded{
We define projects with over 100 commits and 10 stars as ``big'' projects. 
The mean and 3rd quantile values for the number of commits in our dataset are 46 and 12, respectively. 
This aligns with established practices in the literature where thresholds are often set significantly above average to isolate highly active projects.
By setting the threshold at more than double the mean, we ensure that only the top-performing projects are classified as big.
Similarly, the threshold of 10 stars is set based on the mean of 2.33 and 3rd quantile value of 0 for stars.
This indicates that the majority of projects receive few or no stars, reflecting their popularity and community engagement levels.
By selecting projects with at least 10 stars, we focus on those with significant community recognition, capturing less than 1\% of the dataset but representing the most influential projects.
}

\chadded{
The thresholds chosen for ``small'' group, on the other hand, are projects with no stars and fewer than 10 commits to ensure the projects are indeed small and inactive.
This approach ensures that the small group, comprising 62\% of projects, includes those with minimal activity and engagement, consistent with findings that a large proportion of open source projects are relatively inactive~\cite{gousios2012ghtorrent}.
We consider all the other projects that do not fall into either the big or small categories as the ``medium'' group. 
The medium group captures the middle ground, excluding only the extremes, thus providing a balanced representation of the majority of active projects.
}

Using this taxonomy, we counted the number of unique blobs involved in these copy instances between groups.
It should be mentioned that a blob can have several downstream projects that do not necessarily fall into the same group.
Therefore, we considered the biggest downstream project for our analysis purposes.
For example, if a blob originated in a medium project and was reused by both a big and a small project, we count it in the ``medium to big'' category.
\chadded{
Considering the biggest downstream project for each unique blob ensures that the most significant reuse instances are captured.
This approach is supported by research indicating that the impact of code reuse is often determined by the size and activity of the downstream projects utilizing the code~\cite{mockus2007large,weiss1999software}.
By focusing on the largest downstream project, we ensure that our analysis reflects the most substantial and influential reuse cases of a particular blob.
}

\subsubsection{RQ1-c: Do characteristics of the blob affect the probability of reuse?}
The third part of our research question (RQ1) focuses on the properties of reused artifacts.
To address this, we obtained a large random sample of blobs comprising 1/128 of all blobs.
\chadded{
We have to point out that unlike RQ1-b, where we randomly sampled copy instances (meaning all the blobs involved were reused at least once), here we are sampling from the b2tP map that includes all blobs, whether they have been reused or not.
Our dataset is divided into 128 files based on the first two bytes of the blob hash.
Hash functions, by design, distribute input data evenly across the output space.
The use of hash functions to divide data ensures a uniform distribution across the resultant files~\cite{mitzenmacher2017probability}.
By using one of these 128 files as our sample, and given the vast size of the dataset, we ensure that it is an unbiased representation of the entire dataset and that this sample size is sufficient to achieve high statistical power and accuracy in our analyses.
}

We then employed a logistic regression model with the response variable being one for reused blobs and zero for non-reused blobs.
\chadded{
Logistic regression is a robust statistical method used to model the probability of a binary outcome based on one or more predictor variables.
It is widely used in empirical software engineering to understand factors influencing software development practices~\cite{hosmer2013applied}.
By using logistic regression, we can quantify the effect of various predictors on the likelihood of a blob being reused.
}

\chadded{
In this research question, we are concerned with infectiousness based on our SCT framework.
Specifically, we are looking for properties of artifacts that affect their propensity to be reused.
}
\chadded{
The first predictor in our model is the programming language of the blob.
Different programming languages are associated with distinct package managers, development environments, and community cultures, which can influence reuse practices~\cite{mockus2002two}.
For example, the ease of dependency management in languages like Python (via pip) or JavaScript (via NPM) might facilitate reuse more than in languages with less mature package management systems.
Thus, including the programming language as a predictor helps capture these contextual differences.
We anticipate that source code for programming languages such as C, which lack package managers, is likely to be copied more frequently than source code for languages with sophisticated package managers, such as JavaScript.
}

\chadded{
The second predictor is the time of blob creation.
This factor helps account for temporal dynamics by indicating the period during which a blob was created, reflecting different reuse practices over time.
We hypothesize that older blobs were more likely to be reused due to fewer available reusable artifacts in the OSS landscape at the time.
However, the time of creation inherently includes the effect of a blob's availability duration ($t_b(P_d)-t_b(P_o)$), meaning older blobs have had more time to be discovered and reused.
Previous research indicates that the age and visibility of code artifacts influence their reuse~\cite{weiss1999software}.
}

\chadded{
To isolate and examine the influence of the creation period without the confounding effect of longer availability, we introduce the concept of time-limited reuse.
By focusing on copies occurring within specific time intervals after the blob's creation, we remove the advantage of longer visibility and can better assess how the creation period itself influences reuse}\footnote{
This definition is used solely for the purposes of our regression model and subsequent analysis. 
It is not applied in RQ1-a, RQ1-b, or RQ2.}.
\chadded{
We evaluated both one-year and two-year intervals and found similar results.
By evaluating both intervals and finding similar results, we enhance the robustness of our conclusions.
To maintain conciseness and avoid repetition, we report the findings for the two-year interval.
Reporting the two-year interval results provides a balance between sufficient observation time for reuse events and the practical need for concise reporting.
Consequently, we excluded blobs created after May 1, 2020, ensuring that all blobs had at least two years to be potentially reused, providing a consistent time frame for analysis~\cite{weller2016manifesto}.
This approach ensures that our findings are not skewed by varying availability periods.
}

\chadded{
The third predictor is whether the blob is a source code or a binary.
We hypothesize that binaries, identified by their git treatment or file extensions like tar, jpeg, or zip, may exhibit different reuse patterns compared to source code.
We expect that binary files, such as images, might be copied more often because they are easy to understand and reuse but difficult to recreate. 
Unlike other types of files, developers cannot easily extract specific parts or functionalities from binary files.
That is, source code blobs are directly reusable and modifiable, whereas binaries might be reused as-is without modification.
This distinction is important as it affects the ease or necessity of reuse~\cite{gabel2010study}.
Therefore, when it comes to whole-file reuse, which is our definition of reuse in this work, we anticipated that binary blobs are more likely to be copied.
}

\chadded{
The last factor we hypothesize might affect the propensity of a blob to be reused is its size.
The size of a blob can influence its reuse for several reasons.
Larger blobs may contain more functionality, making them more attractive for reuse.
Conversely, smaller blobs may be simpler to integrate into existing projects.
Previous research has indicated that the size of code artifacts can impact their maintainability, comprehensibility, and ultimately their reuse~\cite{capiluppi2003characteristics,mockus2007large}.
}

\chadded{
To investigate whether a difference exists between the sizes of copied and non-copied blobs, we exclude binary blobs from the analysis.
The size of binary blobs is not comparable to the size of source code blobs due to their fundamentally different nature.
Binary blobs often include compiled code, media files, or compressed archives, which do not provide a meaningful comparison to plain text source code in terms of size.
Because of these differences, we did not incorporate blob size as a predictor in our logistic regression model.
Including binary blobs could skew the results and lead to misleading conclusions.
Instead, we perform a t-test to compare the sizes of copied blobs and non-copied blobs.
The t-test is a robust statistical method used to determine whether there is a significant difference between the means of two groups~\cite{student1908probable}.
By applying the t-test, we can rigorously assess whether blob size influences the likelihood of reuse.
}

\subsubsection{RQ1-d: Do characteristics of the originating project affect the probability of reuse?}
The fourth part of RQ1 concerns the chances of finding or being aware of a blob approximated by signals at the project level.
\chadded{This is the exposure factor in the SCT framework.}
To conduct this study, we use WoC's MongoDB project database to randomly sample one million projects\chadded{, comprising nearly 1\% of all projects indexed by WoC, to achieve a balance between statistical validity and computational feasibility.
A sample size of one million is large enough to provide a representative snapshot of the entire population.
}

We then search the reuse instances $(t_b(P_o), t_b(P_d))$ in our Ptb2Pt map to determine if the project originated at least one reused blob.
A logistic regression model with the response variable being one if the project has introduced at least one reused blob (and zero otherwise) is then constructed.
The predictors in the project-level model include the number of commits, blobs, authors, forks, earliest commit time, the activity duration of the project (the time between the first and the last commit in that project)\chadded{, the binary ratio (the ratio of binary blobs to total blobs), and the programming language}.
We also use the number of GitHub stars for each project as a predictor. 
This data in WoC (number of stars) is sourced from GHTorrent~\cite{gousios2013ghtorent}.

\chadded{The choice of these predictors for our model is based on the current literature on relevant project properties.}
\begin{itemize}
    \item \chadded{
    \textbf{Number of commits}: A strong indicator of project activity and maintenance.
    Research shows that projects with higher commit frequencies tend to have more active development and are more likely to be reused due to their perceived reliability and continuous improvement~\cite{koch2002effort}.
    }  
    \item \chadded{
    \textbf{Number of blobs}: Represents the volume of content and potential reusable components. 
    Larger projects with more blobs are likely to offer more opportunities for reuse~\cite{mockus2007large}.
    It can also indicate the project's complexity and modularity.
    Projects with more files may be more modular and provide more reusable components~\cite{bird2009putting}.
    }
    \item \chadded{
    \textbf{Number of authors}: Reflects the collaborative nature of a project.
    Projects with more contributors tend to have diverse expertise and higher code quality, increasing the likelihood of reuse~\cite{crowston2005social}.
    }
    \item \chadded{
    \textbf{Number of forks}: A proxy for the project's popularity and community engagement.
    Projects with more forks are often viewed as valuable and trustworthy, increasing their reuse potential~\cite{tsay2014influence}.
    }
    \item \chadded{
    \textbf{Earliest commit time and the activity duration}: Provide insights into the project's maturity and stability.
    Older and long-active projects are more likely to be well-established and reused~\cite{capiluppi2003characteristics}.
    }
    \item \chadded{
    \textbf{GitHub Stars}: A form of social endorsement, indicating community approval and interest.
    Projects with more stars are likely to be considered high-quality and reliable, making them more attractive for reuse~\cite{borges2016predicting}.
    }
    \item \chadded{
    \textbf{Binary ratio}: Defined as the ratio of binary blobs to total blobs, this can impact the reuse potential of a project.
    Binary blobs, such as compiled code or media files, often indicate pre-packaged functionalities or resources that are ready for use.
    A higher binary ratio may suggest that a project provides ready-to-use components, which can facilitate reuse~\cite{mockus2007large}.
    }
\end{itemize}

Regarding language assignment, at the blob-level, WoC's b2sl map was used for blob language detection based on file extensions.
\chadded{This method is straightforward and effective for identifying the programming languages of individual blobs.}
Nevertheless, assigning a primary language to a project is more complex due to the use of multiple languages in most projects.
WoC's MongoDB project database provides counts of files with each language extension, allowing us to pick the most frequent extension as the project's main language.
For our study, we considered only a subset of blobs, specifically originating blobs (blobs first seen in OSS within the project), and assumed the most common language among these blobs as the project's primary language.
\chadded{This approach aligns with the practice of determining the dominant language based on primary contributions~\cite{vasilescu2015data}.}

\subsection{RQ2: How do developers perceive and engage with copy-based reuse?}\label{m:rq2}
\chadded{The second research question in our study aims to triangulate the quantitative results and understand how developers perceive and engage with copy-based reuse.}
While quantitative research often focuses on metrics such as frequency, intensity, or duration of behavior, qualitative methods are better suited to explore the beliefs, values, and motives underlying these behaviors~\cite{castleberry2018thematic}.

\chadded{
Using a questionnaire for triangulation allows us to obtain self-reported data, which can confirm or challenge the quantitative findings.
This method helps identify any discrepancies and provides a deeper understanding of participant behavior~\cite{denzin2017research}.
In our study, the questionnaire included a direct question (``Did you create or copy this file?'') to gather self-reported data on whether participants copied the blob, offering a direct measure to compare against the quantitative results.
}

\chadded{
Additionally, based on the Social Contagion Theory (SCT) framework, we hypothesize that the characteristics of the destination project and/or author influence reuse activity.
However, treating all reusers the same could be problematic, as developers may have fundamentally different reasons for reuse.
Motivations for reuse can vary widely based on individual needs, project requirements, and perceived benefits from the reused code~\cite{mockus2007large,frakes1995sixteen}.
Our primary focus was to understand these motivations to categorize different types of reuse, potentially providing more insight into measuring susceptibility for future research.
By categorizing motivations, we aim to identify distinct patterns and factors influencing reuse behavior, facilitating the development of targeted strategies to enhance code reuse practices.
This approach aligns with qualitative research methods that seek to explore complex phenomena through detailed, contextualized analysis~\cite{creswell2017research}.
}

To gain insights into the motivations behind copy-based reuse, we conducted an online survey \chadded{targeting both the authors of commits introducing reused blobs and the authors of commits in the originating repositories.
The survey aimed to capture a range of experiences and perceptions related to copy-based reuse}\footnote{
The survey and its procedure was approved by our institutional review board, ensuring that it adhered to ethical guidelines for research involving human subjects.}.

\subsubsection{Survey Content and Questions}
\chadded{
The survey included questions about the nature of the file, why it was needed, how it was chosen, and whether developers would use tools to manage reused files.
General questions about the repositories and developers' expertise were also included.
Notably, the question about the reason for needing the file was open-ended to capture unbiased and detailed responses about the motivations for reuse.
}
All the questions were optional, except for the very first one, which asked if the respondent had created or reused the file.
We chose not to directly ask why did developers choose to copy to avoid provoking legal and ethical concerns about copy-based reuse.
For this reason, instead, we asked: ``Why was this file needed? How did it help your project?''\footnote{
See online appendix for survey questions.}.

\chadded{
Furthermore, we asked developers if the project in which the file resides was intended to be used by other people.
Understanding whether creators intend for their resources to be reused helps assess the cultural and strategic aspects of OSS development.
If a significant portion of creators design their code with reuse in mind, it indicates a collaborative ecosystem where resources are shared and built upon. 
}

\chadded{
We also asked a series of Likert scale (on a scale from 1 to 5) questions as follows.
}

\begin{itemize}
    \item \chadded{
    \textbf{``To what extent did this file help you?''}: Gauging how helpful creators and reusers find the reused blobs provides quantitative data on the perceived value of the reused code.
    Comparing the ratings between creators and reusers highlights any discrepancies or alignment in perceived usefulness.
    }
    \item \chadded{
    \textbf{``To what extent were you concerned about potential bugs in this file?''}: Investigating reusers’ concerns about bugs in reused code sheds light on the perceived risks associated with this practice.
    Understanding the level of concern can indicate how much trust reusers place in the original code’s quality.
    }
    \item \chadded{
    \textbf{``How important is it for you to know if the original file has been changed?''}: Understanding reusers’ concerns about changes in the original files helps identify potential issues related to the stability and continuity of reused code.
    Frequent changes can disrupt the functionality of dependent projects.
    }
    \item \chadded{
    \textbf{``How likely would you use a package manager which could handle changes to this file if there was one?''}: Understanding the likelihood of reusers adopting a package manager if available provides insights into the demand for tools that can streamline and manage code reuse.
    }
\end{itemize}

\subsubsection{Sampling Strategy}
\chadded{
To ensure a representative and comprehensive sample, we stratified the data along several dimensions.
Stratified sampling ensures that all relevant subgroups are adequately represented in the survey, enhancing the generalizability of the findings~\cite{creswell2017research}.
By considering multiple dimensions such as productivity, popularity, copying patterns, file types, and temporal aspects, we ensure a comprehensive analysis that captures the diversity of reuse behaviors in the OSS community:
}
\begin{itemize}
    \item \textbf{Productivity and Popularity}: Based on the number of commits and stars, we differentiated between high and low productivity/popularity projects (similar to RQ1-b).
    \item \textbf{Copying Patterns}: We distinguished between instances where only a few files were copied versus multiple files, as these might indicate different reuse behaviors.
    \item \textbf{File Extension}: We included various file types and programming languages to capture a diverse range of reuse scenarios.
    \item \textbf{Temporal Dimensions}: We considered the blob creation time and the delay from creation to reuse to understand temporal patterns in reuse behavior.
\end{itemize}

\subsubsection{Survey Design}
For each copy instance, we targeted the author of the commit introducing the blob into the destination repository and the author of the commit in the originating repository\footnote{
Only if they had explicitly disclosed their email address on their public profile.}.
\chadded{This dual perspective allowed us to capture both the originator’s and the reuser’s viewpoints, offering a more comprehensive understanding of the reuse dynamics.}

% survey stat
We conducted three rounds of surveys\chadded{, progressively expanding the sample size and refining the questions based on feedback and preliminary results.
We chose to conduct our survey in three steps to ensure a thorough and iterative approach to understanding developer motivations behind copy-based reuse.}

\begin{enumerate}
    \item We handpicked 24 developers (12 creators and 12 reusers) for an initial survey with open-ended questions.
    This round aimed to gather in-depth qualitative data and identify key themes.
    This small, purposive sample size allows for deep, exploratory insights, which are important for the initial stages of qualitative research~\cite{guest2006many}.
    \item The survey was sent to 724 subjects (329 creators and 395 reusers) with a mix of open-ended and multiple-choice questions.
    This round helped validate and refine the themes identified in the first round.
    The increased sample size in this round provides more data to ensure that the themes and patterns observed are not idiosyncratic but rather indicative of broader trends.
    This intermediate sample size balances the need for more extensive data while still allowing for qualitative depth~\cite{mason2010sample}.
    \item The survey was expanded to 8734 subjects (2803 creators and 5931 reusers), with most questions being multiple-choice to facilitate quantitative analysis, except for the open-ended question about the reason for needing the file.
    The large sample size in this final round ensures that the findings are statistically significant and generalizable across the broader population of developers involved in copy-based reuse.
    This sample size aligns with recommendations for achieving sufficient statistical power in survey research~\cite{krejcie1970determining}.
\end{enumerate}

\chadded{
The reason behind the seemingly random numbers of survey subjects in the three rounds is that after sampling our data, we had to perform data cleansing and preparation to reach the survey target audience.
This process normally caused some samples to be removed.
Initially, we chose sample sizes of 30, 1,000, and 10,000 respondents for the three rounds respectively, but after the data cleansing process, the actual numbers were lower.
}

\subsubsection{Thematic Analysis}
The thematic analysis allows us to systematically identify patterns and themes within qualitative data, providing deep insights into the reasons behind copy-based reuse~\cite{braun2006using}.
To analyze the survey responses, we followed a structured thematic analysis process as outlined in~\cite{yin2015qualitative}:

\begin{enumerate}
    \item \textbf{Compiling}: First author compiled all responses.
    \item \textbf{Disassembling}: Each author individually analyzed and coded the responses to identify ideas, concepts, similarities, and differences~\cite{austin2014qualitative,sutton2015qualitative}.
    \item \textbf{Reassembling}: The coded responses were organized into meaningful themes by each author independently, focusing on identifying different types of reuse~\cite{braun2006using}.
    \item \textbf{Interpreting and Concluding}: The authors discussed and compared the themes, clarifying and organizing them to ensure a coherent and comprehensive understanding.
    The final themes were then used to reclassify and interpret all survey responses.
\end{enumerate}

\section{Results \& Discussions}

The numbers presented in this section are derived from version U\footnote{
https://bitbucket.com/swsc/overview} 
of WoC, which was the most recent version available at the time of this analysis.

\subsection{RQ1: How much copy-based reuse occurs? What factors affect the propensity to reuse?}
\subsubsection{RQ1-a: How extensive is copying in the entire OSS landscape?}
We identified nearly 24 billion copy instances (unique tuples containing the blob and originating and destination projects) encompassing more than 1 billion distinct blobs.
With approximately 16 billion blobs in the entire OSS landscape (as approximated by WoC), 6.9\% of the blobs have been reused at least once, and each reused blob is copied to an average of 24 other projects (see Table~\ref{tbl:stat}).

\begin{table}[h!]
\centering
\caption{Basic Statistics of Reuse Instances}
\begin{tabular}{lccc}
  \toprule
  & \textbf{Count} & \textbf{Total} & \textbf{\%} \\ 
  \midrule
    Reuse instances & 23,914,332,270 & - & - \\ 
    Blobs & 1,084,211,945 & 15,698,467,337 & 6.9\% \\ 
    Originating projects & 31,706,416 & 107,936,842 & 29.4\% \\ 
    Destination projects & 86,483,266 & 107,936,842 & 80.1\% \\ 
  \bottomrule
\end{tabular}
\label{tbl:stat}
\end{table}

Nearly 32 million projects (about 30\% of the nearly 108 million deforked OSS projects indexed by WoC) originated at least one reused blob.
Over 86 million projects have copied these blobs, meaning 80\% of OSS projects have reused blobs from another project at least once.

\begin{framed}
\noindent
\colorbox{lightgray}{
\parbox{\dimexpr\linewidth-2\fboxsep-2\fboxrule\relax}{
\color{darkblue}
\noindent\textbf{RQ1-a Key Findings:}
\begin{enumerate}
  \item We identified nearly 24 billion copy instances encompassing more than 1 billion distinct blobs.
  \item 6.9\% of all the blobs in the entire OSS have been reused at least once.
  \item About 30\% of all OSS projects originated at least one reused blob, and 80\% of projects have reused blobs at least once.
\end{enumerate}
}}
\end{framed}

\chadded{
The extensive reuse observed highlights the efficiency gains in OSS development, as projects benefit from existing code to accelerate development cycles and reduce costs.
The widespread reuse also raises significant security concerns, as vulnerabilities in copied code can propagate across numerous projects.
This necessitates improved vulnerability detection and management practices to ensure the integrity of reused code.
Additionally, License violations due to improper code reuse can lead to legal challenges and compliance issues, underscoring the importance of clear licensing and adherence to open source policies.
Furthermore, our identification of blob-level reuse, which only accounts for exact matches and not slight modifications, suggests that the actual extent of code reuse might be even higher.
The findings advocate for the development of better tools and infrastructure to manage copy-based reuse, including automated detection of security and legal risks, and tools for maintaining code quality in reused components.
}

\subsubsection{RQ1-b: Is copy-based reuse limited to a particular group of projects?}
The numbers already demonstrate the prevalence of copy-based reuse in the OSS community.
To understand how this reuse activity is distributed across different groups of projects, we constructed a contingency table as explained in the methods section.
Each blob's originating project is unique and falls into one of three categories (big, medium, and small).
\chadded{However, downstream projects are not unique and we consider the largest downstream project for each blob.}

Our analysis revealed nearly 112 million unique blobs reused in our 640 million sample copy instances, with nearly 13 million of these blobs reused by at least one big project (see Table~\ref{tbl:bcouts}). 
This indicates that more than 11\% of blobs are reused at least once by at least one big project, showing that copy-based reuse is not limited to small projects but is a widespread phenomenon in the OSS community.

\begin{table}[h!]
\centering
\caption{Blob Counts in Reuse Sample}
\begin{tabular}{ccccc|c}
  \toprule
  & & \multicolumn{3}{c}{\textbf{Biggest Downstream Projects}} & \textbf{Total} \\ 
  & & Big & Medium & Small &  \\
  \midrule
  \textbf{Upstream} & Big & 6,748,621 & 22,273,811 & 6,515,122 & 35,537,554 (31.8\%) \\
  \textbf{Projects} & Medium & 5,348,651 & 36,434,732 & 14,552,148 & 56,335,531 (50.3\%) \\
  & Small & 691,644 & 10,151,838 & 9,231,618 & 20,075,100 (17.9\%) \\
  \midrule
  \textbf{Total} & & 12,788,916 (11.4\%) & 68,860,381 (61.5\%) & 30,298,888 (27.1\%) & \textbf{111,948,185} \\
  \bottomrule
\end{tabular}
\label{tbl:bcouts}
\end{table}

However, it is still unclear if these reused blobs are predominantly introduced by big projects.
If this were the case, one could presume that these blobs are mostly of good quality and not error-prone, making the costs of managing and tracking code propagation through such reuse potentially outweigh the benefits.
Sampling copy instances revealed that big projects are responsible for only about 30\% of reused blobs, while the remaining 70\% are introduced by medium and small projects.
Specifically, nearly 18\% of these blobs are introduced by small projects, with the remaining 50\% coming from medium projects.
Furthermore, even for big projects, almost 50\%\footnote{
$(5,348,651 + 691,644) / 12,788,916$}
of the blobs they reuse originate from medium and small projects (see Table~\ref{tbl:bcouts}).
Therefore, it is evident that not only big projects serve as upstream sources for copy-based reuse.
Indeed, many blobs introduced by medium and small projects are being widely reused.

Even if all widely reused blobs were exclusively introduced by big projects, copy-based reuse still requires management for several reasons.
For example, security vulnerabilities may continue to spread even after the main project has fixed the issue~\cite{9794064}.

\begin{framed}
\noindent
\colorbox{lightgray}{
\parbox{\dimexpr\linewidth-2\fboxsep-2\fboxrule\relax}{
\color{darkblue}
\noindent\textbf{RQ1-b Key Findings:}
\begin{enumerate}
  \item 32\% of reused blobs originate from big projects, which comprise 1\% of the total projects.
  \item 18\% of reused blobs originate from small projects, which make up 62\% of the total projects.
  \item 50\% of reused blobs originate from medium projects, which represent 37\% of the total projects.
  \item Nearly 50\% of blobs reused by big projects originate from medium and small projects, highlighting significant cross-category reuse.
\end{enumerate}
}}
\end{framed}

\chadded{
Our findings demonstrate that a significant portion of reused code in the OSS community comes from medium and small projects, challenging the assumption that high-quality code predominantly originates from large projects.
This implies a diverse quality spectrum in reused code and underscores the importance of ensuring quality and security across all project sizes, as vulnerabilities in smaller projects can propagate widely.
Tools that can track the origin and usage of blobs are essential to ensure timely updates and fixes across the OSS ecosystem, mitigating risks associated with vulnerabilities and outdated code.
The widespread nature of code reuse across projects of all sizes, emphasizes the need for quality assurance, effective management, and community collaboration to maintain the health and sustainability of the OSS landscape.
}

\subsubsection{RQ1-c: Do characteristics of the blob affect the probability of reuse?}
In this section, we first demonstrate the reuse trends, followed by the logistic regression model predicting the probability of a blob being reused.
Additionally, we present the reuse propensity per language and show the difference in blob size between reused and non-reused blobs.
Finally, we discuss a case study using JavaScript as an example.

\begin{figure}[h]
     \centering
     \begin{subfigure}[b]{0.495\textwidth}
         \centering
         \includegraphics[width=\textwidth]{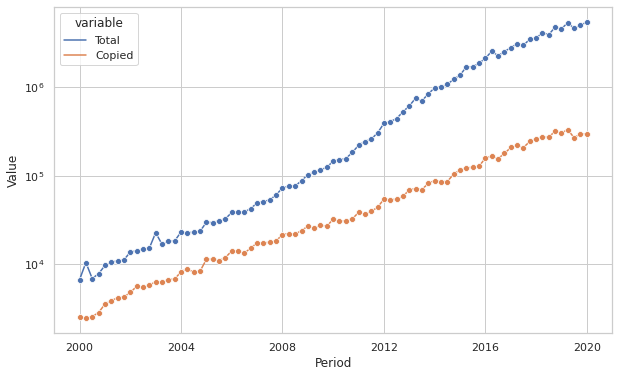}
         \caption{Generated and Reused Blobs Trends}
         \label{fig:time}
     \end{subfigure}
     \hfill
     \begin{subfigure}[b]{0.495\textwidth}
         \centering
         \includegraphics[width=\textwidth]{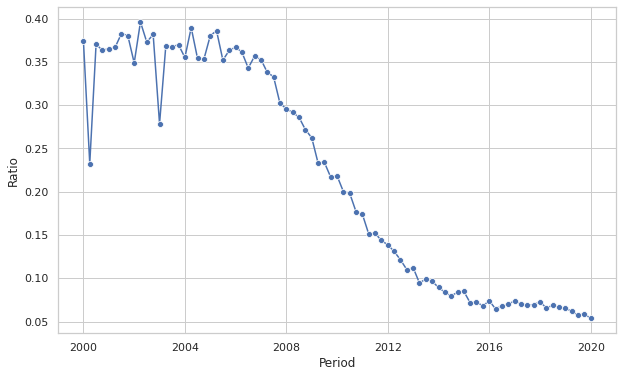}
         \caption{Reused to Generated Blobs Ratio Trend}
         \label{fig:ratio}
     \end{subfigure}
        \caption{Quarterly Reuse Trends}
        \label{fig:three graphs}
\end{figure}

\paragraph{Reuse Trends}
As explained in the methods section, we use a 2-year-limited copying definition in the RQ1-c and RQ1-d models and results.
This means that we consider a blob reused only if it has been reused within 2 years of its creation.
With this definition, 7.5\% of blobs have been reused.
Figure~\ref{fig:time} shows the total counts of new blobs and copied blobs for each quarter since the year 2000\footnote{
The number of projects and blobs was much smaller before 2000.}.
Both counts exhibit rapid growth, although the growth in new blob creation appears to outpace that of copying.
To investigate this difference, Figure~\ref{fig:ratio} shows the reuse propensity measured via the reuse ratio (reused blobs divided by total blobs), confirming that new blob creation has outpaced copied blobs since 2006 when the ratio began to decline.

\paragraph{Logistic Regression Model}
We expect the nature of the blob to affect its propensity to be reused.
To test this hypothesis, we use a logistic regression model where the response variable is set to one if the blob has been copied at least once (i.e., has been committed in at least two projects) within two years of its creation, and zero otherwise.
We used WoC definition of the programming language associated with each blob and categorized less common programming languages in the sample as ``other''.
The descriptive statistics of the variables are presented in Table~\ref{tbl:des_b}.

\begin{table}[h!]
\centering
\caption{Blob-level Model - Descriptive Statistics}
\begin{tabular}{lcccc}
  \toprule
  \textbf{Variable} &  & \textbf{Statistics} & & \\ 
  \midrule
  \midrule
  Reused & Yes: 6,419,388 (7.5\%) & & No: 78,136,705 (92.5\%) & \\ 
  \midrule
  Language & JavaScript & Java & C & (Other) \\ 
  (Counts) & 11,122,849 & 4,579,458 & 3,460,733 & 65,393,053 \\ 
  \multicolumn{5}{c}{\dotfill} \\
  Creation Time & 5\% & Median & Mean & 95\% \\ 
  %Time & 1343565173 & 1517991355 & 1495982603 & 1582853246 \\ 
  (Date) & 7/29/2012 & 2/7/2018 & 5/28/2017 & 2/28/2020 \\ 
  \multicolumn{5}{c}{\dotfill} \\
  Binary & Yes: 18,516,721 (21.8\%) & & No: 66,039,372 (78.2\%) & \\ 
  \bottomrule
\end{tabular}
\label{tbl:des_b}
\end{table}

\chadded{
The sample dataset is predominantly composed of blobs written in JavaScript, with significant counts also in Java and C.
Additionally, the distribution of blob creation time is provided, showing a median date of February 7, 2018.
Furthermore, a notable proportion of the blobs, 21.8\%, are binary.
}

\chadded{The results of our logistic regression model are shown in Tables~\ref{tbl:model_b} and~\ref{tbl:anova_b}.}
The model shows that the coefficients for all predictors are statistically significant with p-values less than 0.0001, meaning they impact the probability of a blob being reused (see Table~\ref{tbl:model_b}). 

\begin{table}[h]
\centering
\caption{Blob-level Model - Coefficients}
\begin{tabular}{lrrrr}  
  \hline
   & \textbf{Estimate} & \textbf{Std. Error} & \textbf{z value} & \textbf{Pr($>$$|$z$|$)} \\ 
  \hline
  (Intercept) & -18.0293 & 0.0186 & -967.07 & \cellcolor{lightgreen}$<2\times10^{-16}$ \\ 
  Binary & 0.4775 & 0.0010 & 460.16 & \cellcolor{lightgreen}$<2\times10^{-16}$ \\ 
  Creation Time & 0.8108 & 0.0010 & 828.34 & \cellcolor{lightgreen}$<2\times10^{-16}$ \\ 
  C & 0.7142 & 0.0017 & 426.32 & \cellcolor{lightgreen}$<2\times10^{-16}$ \\ 
  C\# & -0.1277 & 0.0033 & -38.15 & \cellcolor{lightgreen}$<2\times10^{-16}$ \\ 
  Go & 0.3095 & 0.0065 & 47.74 & \cellcolor{lightgreen}$<2\times10^{-16}$ \\ 
  JavaScript & -0.0832 & 0.0015 & -56.21 & \cellcolor{lightgreen}$<2\times10^{-16}$ \\ 
  Kotlin & -0.5606 & 0.0133 & -42.02 & \cellcolor{lightgreen}$<2\times10^{-16}$ \\ 
  ObjectiveC & 0.0810 & 0.0066 & 12.30 & \cellcolor{lightgreen}$<2\times10^{-16}$ \\ 
  Python & -0.0327 & 0.0030 & -10.97 & \cellcolor{lightgreen}$<2\times10^{-16}$ \\ 
  R & 0.4070 & 0.0083 & 49.22 & \cellcolor{lightgreen}$<2\times10^{-16}$ \\ 
  Rust & 0.0879 & 0.0095 & 9.30 & \cellcolor{lightgreen}$<2\times10^{-16}$ \\ 
  Scala & -0.6168 & 0.0123 & -50.21 & \cellcolor{lightgreen}$<2\times10^{-16}$ \\ 
  TypeScript & 0.1827 & 0.0046 & 39.38 & \cellcolor{lightgreen}$<2\times10^{-16}$ \\ 
  Java & 0.0794 & 0.0019 & 42.37 & \cellcolor{lightgreen}$<2\times10^{-16}$ \\ 
  PHP & 0.3561 & 0.0024 & 151.14 & \cellcolor{lightgreen}$<2\times10^{-16}$ \\ 
  Perl & 0.7664 & 0.0082 & 92.95 & \cellcolor{lightgreen}$<2\times10^{-16}$ \\ 
  Ruby & -0.4782 & 0.0044 & -108.58 & \cellcolor{lightgreen}$<2\times10^{-16}$ \\ 
  \hline
\end{tabular}
\label{tbl:model_b}
\end{table}

The ANOVA table (Table~\ref{tbl:anova_b}) provides insights into the significance of different variables.
We see that all the predictors have p-value equal to zero, meaning that the null hypothesis\footnote{
H0: The reduced model (without the predictor) provides a fit to the data that is not significantly worse than the full model (with the predictor). 
This suggests that the predictor does not significantly improve the model's fit.
} can be rejected.
The null deviance is 45,438,151, which represents the deviance of a model with only the intercept.
Adding the Binary variable reduces the deviance by 124,114, indicating its strong influence on reuse likelihood.
The Creation Time variable further reduces the deviance by 830,322, highlighting its importance in predicting reuse.
The ``Language'' variable also reduces the deviance by 230,614.
Although these reductions might seem small relative to the null deviance, they are statistically significant given the large sample size and the high degrees of freedom involved.

\begin{table}[h!]
\centering
\caption{Blob-level Model - ANOVA Table}
\begin{tabular}{lrrrrr}
  \toprule
  & \textbf{Df} & \textbf{Deviance} & \textbf{Resid. Df} & \textbf{Resid. Dev} & \textbf{p.value} \\ 
  \midrule
  NULL &  &  & 84,556,092 & 45,438,151.00 &  \\ 
  Binary & 1 & 124,114.20 & 84,556,091 & 45,314,036.80 & \cellcolor{lightgreen}$<2\times10^{-16}$ \\ 
  Creation Time & 1 & 830,322.63 & 84,556,090 & 44,483,714.17 & \cellcolor{lightgreen}$<2\times10^{-16}$ \\ 
  Language & 15 & 230,614.17 & 84,556,075 & 44,253,100.00 & \cellcolor{lightgreen}$<2\times10^{-16}$ \\ 
   \bottomrule
\end{tabular}
\label{tbl:anova_b}
\end{table}

\begin{figure}[h]
    \centering
    \includegraphics[width=\linewidth]{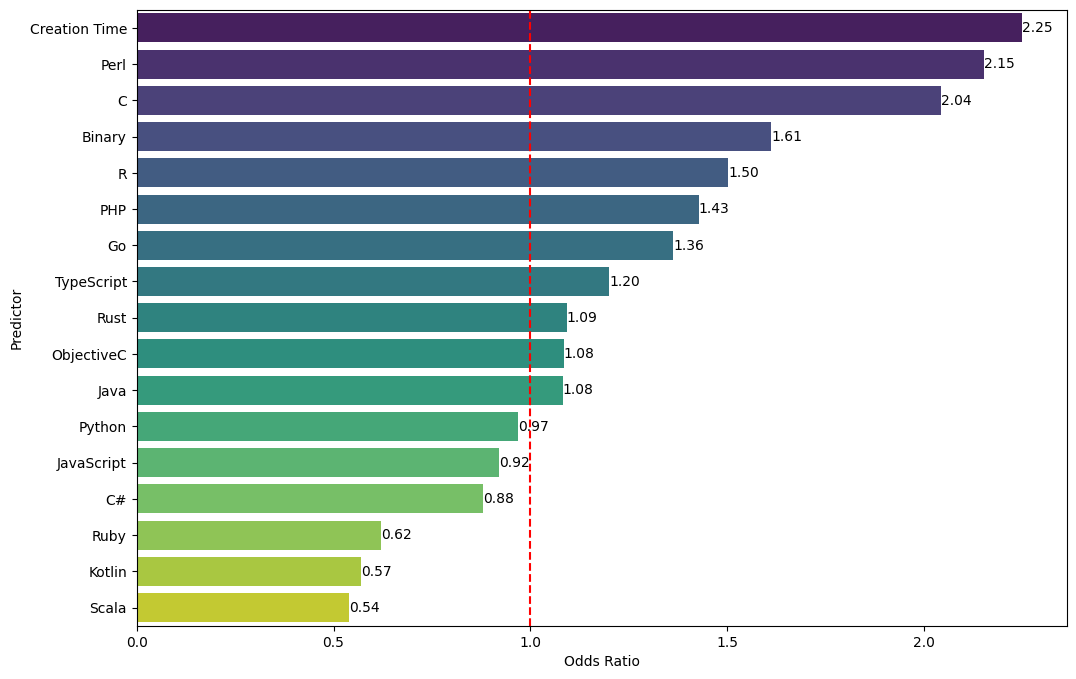}
    \caption{Blob-level Model - Logistic Regression Odds Ratios}
    \label{fig:blob-odds}
\end{figure}

\chadded{
To assess the direction and the size of predictor effects, we need to go further.
In a logistic regression model, a positive coefficient estimate indicates that as the predictor variable increases, the odds of the outcome occurring increase, while a negative coefficient estimate indicates that as the predictor variable increases, the odds of the outcome occurring decrease.
Since the coefficients represent the change in the log-odds of the outcome for a one-unit increase in the predictor, we transform these coefficients to odds ratios by exponentiating them to interpret the actual impact of each predictor.
The odds ratio indicates how the odds of the outcome change with a one-unit increase in the predictor.
The results are shown in Figure~\ref{fig:blob-odds}.
This graph displays the odds ratios for various predictors in the logistic regression model at the blob level.
An odds ratio greater than 1 indicates an increase in the likelihood of reuse, while an odds ratio less than 1 indicates a decrease.
}

\chadded{
The creation time has the highest positive coefficient.
The time variable in the model represents the time elapsed from the blob's creation until current time, meaning that older blobs have higher time values.
The positive coefficient indicates that newer blobs (with smaller time values) are less likely to be reused.
This is not because they have been visible for a shorter duration (as we controlled for this with the time-bound definition of reuse), but likely due to other factors we hypothesized, such as fewer artifacts being available for reuse at the time of their creation.
}

\chadded{
Binary blobs show a significant increase in reuse likelihood with an odds ratio of 1.63.
Given this confirmed effect, we calculated the reuse propensity for binary and non-binary blobs separately.
The results showed that 9.5\% of binary blobs were reused, compared to 7.0\% of non-binary blobs in our sample.
}

\chadded{
Different programming languages show varied impacts on reuse likelihood.
Blobs written in Perl, C, R, PHP, Go, TypeScript, Objective-C, Java, and Rust are more likely to be reused, with Perl showing the highest odds ratio.
In contrast, blobs written in Kotlin, Scala, Ruby, C\#, JavaScript, and Python are less likely to be reused, with Kotlin and Scala showing the most significant negative coefficients.
This variability suggests that certain languages, perhaps due to their prevalence or specific use cases, are more conducive to code reuse.
}

\paragraph{Per-Language Propensity}
\chadded{
Following our logistic regression results, which demonstrated that programming language is a statistically significant factor in reuse probability of a blob, we calculated the propensity to copy for each programming language, measured as the percentage of reused blobs within that language (see Table~\ref{tbl:prop}).
The results show that blobs written in Perl have the highest propensity to be reused at 18.5\%, indicating a strong tendency for code reuse among Perl developers.
Conversely, Kotlin has the lowest propensity at 3.0\%, suggesting minimal code reuse in this language.
Languages such as C (15.2\%) and PHP (9.9\%) also show high reuse rates, while Python (6.4\%), JavaScript (5.5\%), and TypeScript (6.3\%) have lower rates.
Other languages like Java (7.8\%), Go (7.9\%), and R (9.8\%) fall in the middle range, with moderate reuse rates.
}

\begin{table}[h]
\centering
\caption{Blob-level - Propensity to Reuse}
\begin{tabular}{lc|lc|lc}
  \hline
  \textbf{Language} & \textbf{Ratio} & \textbf{Language} & \textbf{Ratio} & \textbf{Language} & \textbf{Ratio} \\ 
  \hline
  C & 15.2\% & ObjectiveC & 8.4\% & TypeScript & 6.3\% \\ 
  C\# & 6.0\% & Python & 6.4\% & Java & 7.8\% \\ 
  Go & 7.9\% & R & 9.8\% & PHP & 9.9\% \\ 
  JavaScript & 5.5\% & Rust & 6.7\% & Perl & 18.5\% \\ 
  Kotlin & 3.0\% & Scala & 3.8\% & Ruby & 5.1\% \\ 
   \hline
\end{tabular}
\label{tbl:prop}
\end{table}

\begin{figure}[h]
    \centering
    \includegraphics[width=\linewidth]{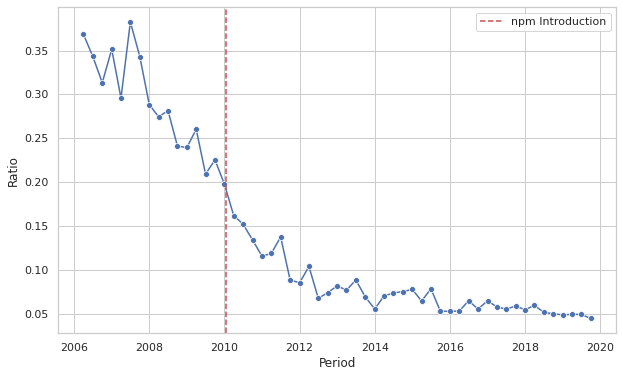}
    \caption{Reused Blobs to Total Generated Blobs Ratio Trend in JavaScript}
    \label{fig:js}
\end{figure}

% JS
\paragraph{JavaScript Example}
\chadded{
The significant role of programming language in reuse activity might have several underlying reasons, as previously discussed.
One such reason is the presence of a reliable package manager.
If true, substantial improvements in a package manager should reduce the propensity to reuse an artifact.
To examine this, we analyzed the timeline of the reuse ratio for JavaScript, shown in Figure~\ref{fig:js}.
The figure indicates a sharper decrease in the slope around 2010, the year the NPM package manager was introduced.
This downward trend continues until mid-2013, when the copying activity rate drops to around 7\% and then levels off.
This pattern supports the hypothesis that the introduction and adoption of NPM significantly reduced code reuse through copying.
}

\chadded{
However, it is important to note that this is just an illustration, and further research is needed to understand this phenomenon fully.
Our current study was not focused on this aspect, so we did not conduct an in-depth analysis.
Additional investigations with more data points and comparisons with other languages that have introduced similar improvements in their package management systems are necessary to confirm that the observed effect is not coincidental or specific to JavaScript alone.
}

% size
\paragraph{Blob Size}
\chadded{
The final predictor we hypothesized to affect the reuse probability of a blob was its size.
To investigate whether there is a significant difference between the sizes of copied and non-copied blobs, we conducted a t-test comparing these sizes.
Our analysis revealed a significant difference (p-value < 2.2e-16), indicating that, on average, copied blobs are smaller than non-copied blobs.
}

\chadded{
However, the effect varies by language.
Specifically, per-language t-tests reveal that copied blobs are smaller in languages like JavaScript and TypeScript, larger in languages such as C and Python, and remain unchanged in Objective-C, as detailed in Table~\ref{tbl:size}.
For example, in JavaScript, the t-value is -59.9, suggesting that copied blobs are significantly smaller, while in C, the t-value is 195.9, indicating that copied blobs are larger.
Similar patterns are observed in other languages, with TypeScript showing a t-value of -35.9 (smaller copied blobs) and Python a t-value of -5.8 (also smaller copied blobs).
Conversely, languages like Java (t-value 120.7) and PHP (t-value 28.6) show that copied blobs tend to be larger.
}

\begin{table}[h]
\centering
\captionsetup{justification=centering}
\caption{
Size Difference between Reused and non-Reused Blobs \\
(Positive t value means larger reused blobs.)
}
\begin{tabular}{lcc|lcc}
  \hline
  \textbf{Language} & \textbf{t value} & \textbf{p-value} & \textbf{Language} & \textbf{t value} & \textbf{p-value} \\ 
  \hline
  C & 195.9 & \cellcolor{lightgreen}$<2\times10^{-16}$ & Rust & -7.8 & \cellcolor{lightgreen}$<2\times10^{-16}$  \\ 
  C\# & 12.5 & \cellcolor{lightgreen}$<2\times10^{-16}$ & Scala & 9.1 & \cellcolor{lightgreen}$<2\times10^{-16}$ \\ 
  Go & 15.5 & \cellcolor{lightgreen}$<2\times10^{-16}$ & TypeScript & -35.9 & \cellcolor{lightgreen}$<2\times10^{-16}$ \\ 
  JavaScript & -59.9 & \cellcolor{lightgreen}$<2\times10^{-16}$ & Java & 120.7 & \cellcolor{lightgreen}$<2\times10^{-16}$ \\ 
  Kotlin & -14.5 & \cellcolor{lightgreen}$<2\times10^{-16}$ & PHP & 28.6 & \cellcolor{lightgreen}$<2\times10^{-16}$ \\ 
  ObjectiveC & 0.7 & $0.430298$ & Perl & 5.8 & \cellcolor{lightgreen}$<2\times10^{-16}$ \\ 
  Python & -5.8 & \cellcolor{lightgreen}$<2\times10^{-16}$ & Ruby & -24.9 & \cellcolor{lightgreen}$<2\times10^{-16}$ \\ 
  R & -7.6 & \cellcolor{lightgreen}$<2\times10^{-16}$ & Other & -364.9 & \cellcolor{lightgreen}$<2\times10^{-16}$ \\ 
   \hline
\end{tabular}
\label{tbl:size}
\end{table}

\chadded{
This variation highlights that the relationship between blob size and reuse propensity is complex and influenced by language-specific factors.
While our findings demonstrate a general trend of smaller copied blobs, the differing patterns across languages suggest that other underlying factors may be at play.
}

\begin{framed}
\noindent
\colorbox{lightgray}{
\parbox{\dimexpr\linewidth-2\fboxsep-2\fboxrule\relax}{
\color{darkblue}
\noindent\textbf{RQ1-c Key Findings:}
\begin{enumerate}
  \item The reuse ratio is decreasing over time.
  \item 7.5\% of blobs have been reused within two years of creation.
  \item Older blobs, when controlling for the confounding effect of increased visibility, are more likely to be reused.
  \item Binary blobs are 63\% more likely to be reused.
  \item Programming languages significantly impact reuse likelihood.
  Blobs written in languages like Perl, C, R, PHP, Go, TypeScript, Objective-C, Java, and Rust are more likely to be reused, while those written in Kotlin, Scala, Ruby, C\#, JavaScript, and Python are less likely to be reused.
  \item The reuse ratio timeline for JavaScript shows a notable decrease in slope around the year the NPM package manager was introduced.
  \item Copied blobs are generally smaller than non-copied blobs, but this is not consistent across different languages.
  The size difference varies by language, with reused blobs in C, Java, PHP, Go, C\#, Scala, Perl, and Objective-C being larger than non-reused blobs, while in JavaScript, TypeScript, Ruby, Kotlin, Rust, R, and Python, the reused blobs are smaller than non-reused blobs.
\end{enumerate}
}}
\end{framed}

\chadded{
The higher reuse propensity among binary blobs suggests that binaries are inherently more reusable, likely due to their compiled nature, which allows easy integration across projects. 
The lower reuse likelihood of newer blobs indicates a potential issue with the integration and acceptance of recent contributions, possibly due to rapid technological advancements and shifts in development practices.
The significant impact of programming languages on reuse likelihood highlights the importance of language-specific tools and ecosystems.
Languages with higher reuse rates, such as Perl and C, benefit from mature ecosystems, while newer or niche languages like Kotlin and Scala show lower reuse rates, potentially due to smaller communities.
The decline in JavaScript code reuse post-NPM introduction suggests that improved package management can reduce the need for direct code copying, promoting more modular and maintainable codebases.
}

\chadded{
Regarding blob size, the general trend indicates that smaller code artifacts are more reusable, likely due to their simplicity and ease of integration.
However, this trend varies significantly across different programming languages.
For example, in languages like JavaScript and TypeScript, copied blobs tend to be smaller, supporting the idea of writing concise and modular code to enhance reusability.
In contrast, in languages like C and Python, copied blobs are often larger, suggesting that the nature and use cases of these languages might necessitate larger reusable components.
This variation underscores the importance of understanding language-specific factors when considering code reuse management strategies.
}

\subsubsection{RQ1-d: Do characteristics of the originating project affect the probability of reuse?}
In this section, we first present the logistic regression model.
We then demonstrate the per-language reuse propensity and compare it to blob-level results.
Finally, we analyze binary blob reuse.

\paragraph{Logistic Regression Model}
We applied a logistic regression model to determine the likelihood of a project introducing at least one reused blob.
The response variable is binary: 1 if the project has introduced a reused blob, 0 otherwise.
Descriptive statistics for the model variables are presented in Table~\ref{tbl:des_p}.
Consistent with blob-level data, the most frequent languages in our sample are JavaScript and Java.

\begin{table}[h]
\centering
\caption{Project-level Model - Descriptive Statistics}
\begin{tabular}{llcccc}
  \hline
  \textbf{Variable} & \textbf{Description} & & \textbf{Statistics} & & \\ 
  \hline
  \hline
  Reused & Project has at least 1 reused blob & Yes: 205,140 (33.7\%) & & No: 403,195 (66.3\%) & \\ 
  \hline
   & & 5\% & Median & Mean & 95\% \\ 
  Blobs & Number of generated blobs & 1 & 15 & 162.7 & 397 \\ 
  Binary & Binary blobs to total blobs ratio & 0 & 0 & 0.1 & 0.6 \\ 
  Commits & Number of commit & 1 & 5 & 57.0 & 84 \\ 
  Authors & Number of authors & 1 & 1 & 2.5 & 3 \\ 
  Forks & Number of forks & 0 & 0 & 1.5 & 1 \\ 
  Stars & Number of GitHub stars & 0 & 0 & 3.4 & 2 \\ 
  % 1374186230 1522055200 1505443609 1583225199
  Time & Earliest commit time & 7/18/2013 & 3/26/2018 & 9/15/2017 & 3/3/2020 \\ 
  Activity & Total months project was active & 1 & 1 & 2.5 & 8 \\ 
  \multicolumn{6}{c}{\dotfill} \\
  Language& JavaScript \hspace{1cm} Java & Python & PHP & C & (Other) \\ 
  (Counts) & 86,065 \hspace{1.3cm} 43,172 & 40,503 & 24,659 & 22,258 & 391,678 \\ 
  \hline
\end{tabular}
\label{tbl:des_p}
\end{table}

\chadded{
Spearman's correlation analysis, suitable for the observed heavily skewed distributions, is presented in Table~\ref{tbl:cors}.
The number of commits shows a high correlation with two other predictors: activity time (0.68) and the number of blobs (0.67).
These high correlations indicate redundancy, as the number of commits does not add significant information beyond what is already captured by activity time and the number of blobs.
This redundancy can lead to multicollinearity, potentially distorting the model's coefficients and reducing interpretability.
Consequently, we remove the number of commits from the model, simplifying it without sacrificing explanatory power.
All other correlations are below 0.52, which are not concerning.
}

\begin{table}[h]
\centering
\caption{Project-level Model - Spearman's Correlations Between Predictors}
\begin{tabular}{lcccccccc}
  \hline
 & \textbf{Blobs} & \textbf{Binary} & \textbf{Commits} & \textbf{Authors} & \textbf{Forks} & \textbf{Stars} & \textbf{Time} & \textbf{Activity} \\ 
  \hline
  Blobs & 1.00 & 0.46 & \cellcolor{lightred}0.67 & 0.34 & 0.22 & 0.22 & 0.09 & \cellcolor{lightyellow}0.52 \\ 
  Binary & - & 1.00 & 0.18 & 0.12 & 0.06 & 0.05 & 0.02 & 0.14 \\ 
  Commits & - & - & 1.00 & 0.45 & 0.27 & 0.26 & 0.05 & \cellcolor{lightred}0.68 \\ 
  Authors & - & - & - & 1.00 & 0.32 & 0.22 & 0.05 & 0.38 \\ 
  Forks & - & - & - & - & 1.00 & 0.48 & 0.14 & 0.28 \\ 
  Stars & - & - & - & - & - & 1.00 & 0.13 & 0.28 \\ 
  Time & - & - & - & - & - & - & 1.00 & 0.05 \\ 
  Activity & - & - & - & - & - & - & - & 1.00 \\ 
   \hline
\end{tabular}
\label{tbl:cors}
\end{table}

The results for the project-level logistic regression model are shown in Tables~\ref{tbl:model_p} and \ref{tbl:anova_p}.
\chadded{
All the variables in the model have p-values less than 0.05, indicating that they are statistically significant in predicting the likelihood of a project introducing reused blobs (see Table~\ref{tbl:model_p}).
This demonstrates strong evidence against the null hypothesis, suggesting that these variables do have an effect on reuse.
}

\begin{table}[h]
\centering
\caption{Project-level Model - Coefficients}
\begin{tabular}{lcccc}
  \hline
  & \textbf{Estimate} & \textbf{Std. Error} & \textbf{z value} & \textbf{Pr($>$$|$z$|$)} \\ 
  \hline
  (Intercept) & -4.79 & 0.16 & -30.01 & \cellcolor{lightgreen}$<2\times10^{-16}$ \\ 
  Blobs & 0.61 & 0.00 & 228.94 & \cellcolor{lightgreen}$<2\times10^{-16}$ \\ 
  Binary & 0.77 & 0.02 & 40.09 & \cellcolor{lightgreen}$<2\times10^{-16}$ \\ 
  Authors & 0.09 & 0.01 & 8.24 & \cellcolor{lightgreen}$<2\times10^{-16}$ \\ 
  Forks & 0.31 & 0.01 & 27.72 & \cellcolor{lightgreen}$<2\times10^{-16}$ \\ 
  Stars & 0.06 & 0.01 & 7.19 & \cellcolor{lightgreen}$6.61\times10^{-13}$
 \\ 
  Time & 0.10 & 0.01 & 12.00 & \cellcolor{lightgreen}$<2\times10^{-16}$ \\ 
  Activity & 0.07 & 0.01 & 10.48 & \cellcolor{lightgreen}$<2\times10^{-16}$ \\ 
  C & -0.33 & 0.02 & -19.60 & \cellcolor{lightgreen}$<2\times10^{-16}$ \\ 
  C\# & -0.30 & 0.02 & -15.74 & \cellcolor{lightgreen}$<2\times10^{-16}$ \\ 
  Go & -0.29 & 0.04 & -7.70 & \cellcolor{lightgreen}$1.33\times10^{-14}$ \\ 
  JavaScript & 0.21 & 0.01 & 22.58 & \cellcolor{lightgreen}$<2\times10^{-16}$ \\ 
  Kotlin & -0.23 & 0.05 & -4.30 & \cellcolor{lightgreen}$1.75\times10^{-5}$ \\ 
  ObjectiveC & -0.13 & 0.03 & -3.63 & \cellcolor{lightgreen}$0.000288$ \\ 
  Python & -0.19 & 0.01 & -14.78 & \cellcolor{lightgreen}$<2\times10^{-16}$ \\ 
  R & -0.27 & 0.05 & -5.93 & \cellcolor{lightgreen}$3.04\times10^{-9}$ \\ 
  Rust & -0.48 & 0.07 & -6.65 & \cellcolor{lightgreen}$2.87\times10^{-11}$ \\ 
  Scala & -0.27 & 0.07 & -3.79 & \cellcolor{lightgreen}$0.000153$ \\ 
  TypeScript & 0.88 & 0.03 & 34.57 & \cellcolor{lightgreen}$<2\times10^{-16}$ \\ 
  Java & -0.25 & 0.01 & -20.90 & \cellcolor{lightgreen}$<2\times10^{-16}$ \\ 
  PHP & 0.29 & 0.01 & 19.59 & \cellcolor{lightgreen}$<2\times10^{-16}$ \\ 
  Perl & -0.31 & 0.10 & -3.20 & \cellcolor{lightgreen}$0.001395$ \\ 
  Ruby & 0.63 & 0.02 & 33.18 & \cellcolor{lightgreen}$<2\times10^{-16}$ \\ 
   \hline
\end{tabular}
\label{tbl:model_p}
\end{table}

Examining the ANOVA results (Table~\ref{tbl:anova_p}) provides further insight into the impact and significance of these predictors.
We see that all the predictors have p-value equal to zero, meaning that the null hypothesis can be rejected.
The deviance values in the ANOVA table indicate the reduction in model deviance when each predictor is included.
For example, adding the number of blobs to the model reduces the deviance by 131,219.53, a substantial reduction that underscores its critical role in the model.
These results confirm the importance of these predictors in explaining the variability in the likelihood of reuse.

\begin{table}[h!]
\centering
\caption{Project-level Model - ANOVA Table}
\begin{tabular}{lccccc}
  \hline
  & \textbf{Df} & \textbf{Deviance} & \textbf{Resid. Df} & \textbf{Resid. Dev} & \textbf{p.value} \\ 
  \hline
  NULL &  &  & 608,334 & 777,660.48 & \\ 
  Blobs & 1 & 131,219.53 & 608,333 & 646,440.95 & \cellcolor{lightgreen}$<2\times10^{-16}$ \\ 
  Binary & 1 & 662.94 & 608,332 & 645,778.01 & \cellcolor{lightgreen}$<2\times10^{-16}$ \\ 
  Authors & 1 & 926.69 & 608,331 & 644,851.32 & \cellcolor{lightgreen}$<2\times10^{-16}$ \\ 
  Forks & 1 & 2,084.02 & 608,330 & 642,767.30 & \cellcolor{lightgreen}$<2\times10^{-16}$ \\ 
  Stars & 1 & 63.77 & 608,329 & 642,703.53 & \cellcolor{lightgreen}$1.44\times10^{-15}$ \\ 
  Time & 1 & 156.98 & 608,328 & 642,546.54 & \cellcolor{lightgreen}$<2\times10^{-16}$ \\ 
  Activity & 1 & 139.31 & 608,327 & 642,407.24 & \cellcolor{lightgreen}$<2\times10^{-16}$ \\ 
  Language & 15 & 5,178.20 & 608,312 & 637,229.03 & \cellcolor{lightgreen}$<2\times10^{-16}$ \\ 
   \hline
\end{tabular}
\label{tbl:anova_p}
\end{table}

\begin{figure}[h]
    \centering
    \includegraphics[width=\linewidth]{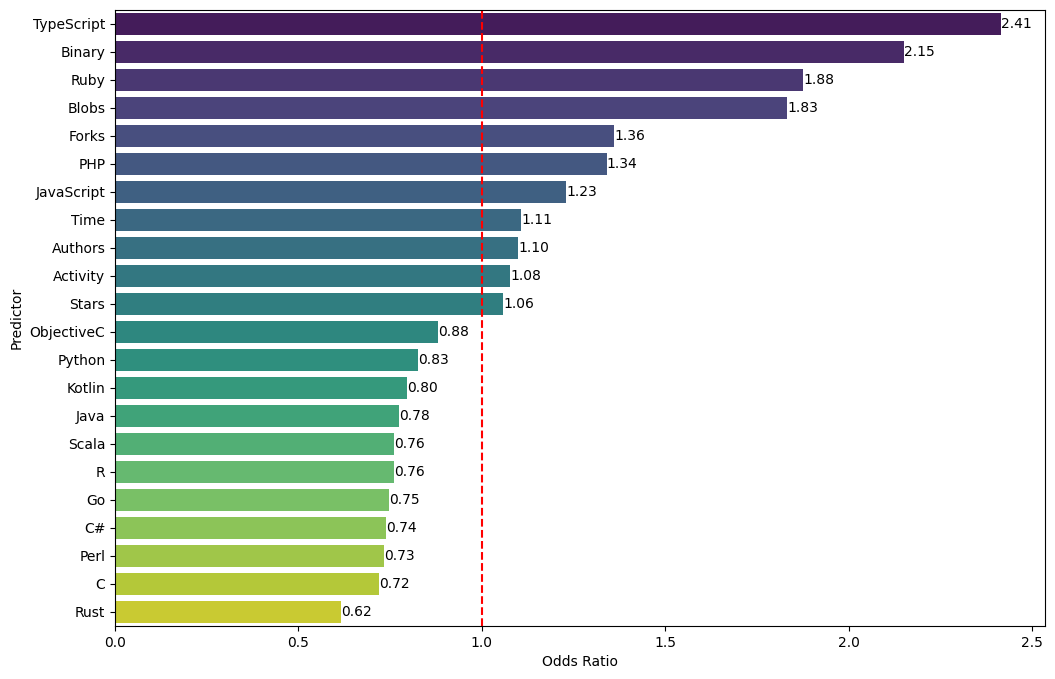}
    \caption{Project-level Model - Logistic Regression Odds Ratios}
    \label{fig:project-odds}
\end{figure}

\chadded{
To understand the size and direction of the impacts, we look at the odds ratios inferred from the logistic regression coefficients.
The odds ratio is calculated as the exponential of the coefficient.
An odds ratio greater than 1 indicates a positive impact, while an odds ratio less than 1 indicates a negative impact.
The results are shown in Figure~\ref{fig:project-odds}.
}

\chadded{
The logistic regression analysis shows that several predictors significantly impact the likelihood of a project having a reused blob.
TypeScript, Binary, Ruby, and Blobs have the strongest positive effects, indicating that increases in these variables substantially raise the odds of a project being reused.
Other positive predictors include Forks, PHP, JavaScript, Time, Authors, Activity, and Stars, which also increase the likelihood, though to a lesser extent.
Conversely, predictors like Rust, C, Perl, C\#, Go, Scala, R, Java, Kotlin, Python, and Objective-C negatively impact the odds, suggesting that increases in these variables decrease the likelihood of a project introducing a reused blob.
}

\chadded{
When interpreting the time variable, it's important to note that since the earliest commit timestamp is represented as a number, we calculated the time elapsed from the earliest commit to the current date for better interpretability.
A larger time value indicates an older earliest commit.
The model shows that time has a positive coefficient, suggesting that the older the earliest commit, the higher the probability of introducing reused blobs.
This result could be influenced by two factors.
First, at the blob-level model, we already observed that older blobs have a higher probability of being reused.
Additionally, while the time-bound definition of reuse controls for the confounding effect of longer visibility at the blob level, it does not account for the longer visibility of the project itself.
Therefore, the observed result might also be affected by the project's age, which implies longer visibility, even though the blob is reused within two years of its creation.
}

\paragraph{Per-Language Propensity}
The project-level model highlights the significance of programming languages in the likelihood of a project introducing a reused blob.
To explore this further, we calculated the percentage of projects in each language that have introduced reused blobs.
From our previous analysis (RQ1-a), we know that approximately 29\% of projects introduced at least one reused blob.
When using the time-bound definition of copying, this ratio increased to 33\% in our sample.
The results for each language are shown in Table \ref{tbl:pl}.

\begin{table}[h]
\centering
\caption{Percentage of Projects Introducing at Least One Reused Blob}
\begin{tabular}{lc|lc|lc}
  \hline
  \textbf{Languages} & \textbf{Ratio} & \textbf{Language} & \textbf{Ratio} & \textbf{Language} & \textbf{Ratio} \\ 
  \hline
  C & 33.2\% & ObjectiveC & 40.0\% & TypeScript & 62.3\% \\  
  C\# & 37.0\% & Python & 30.5\% & Java & 36.2\% \\ 
  Go & 31.3\% & R & 28.5\% & PHP & 46.4\% \\ 
  JavaScript & 41.2\% & Rust & 31.5\% & Perl & 29.9\% \\ 
  Kotlin & 40.0\% & Scala & 36.0\% & Ruby & 51.2\% \\ 
   \hline
\end{tabular}
\label{tbl:pl}
\end{table}

The ratio of projects that have introduced reused blobs varies significantly across different programming languages, offering new insights compared to the blob-level analysis.
For example, projects dominated by TypeScript have the highest probability (62\%) of introducing at least one reused blob.
This finding is particularly interesting because, at the blob level, the propensity to copy in TypeScript was lower than average.
This discrepancy suggests that TypeScript projects, acting as upstream in the language's supply chain, are less centralized.
Developers in this language seem more inclined to incorporate code from various, possibly unknown, projects.

\chadded{
Other languages also show distinct patterns.
For instance, Ruby projects have a high probability (51\%) of reusing blobs, whereas Python projects have a lower probability (30.5\%).
This variation indicates that the likelihood of code reuse is strongly influenced by the primary language of the project, reflecting different practices and community norms across languages.
These insights emphasize the importance of considering programming language when studying code reuse patterns in software projects.
}

To ensure these results are comparable to blob-level analysis, we calculated the copied blob ratio (copied blobs to total blobs) for each project and took the average of this ratio for projects in each language.
An important difference here with the blob-level propensity is that at the blob level, language assignment was based on the file extension of each blob, with binary blobs categorized as ``Other''.
In this project-level analysis, the language of a blob is determined by the predominant language of the project it belongs to.
For example, a Python-written blob in a C-dominated project is counted as a C blob.
Similarly, binary blobs are assigned the language of the dominant language in their respective projects.
The results of this new definition are shown in Table \ref{tbl:cr}.

\begin{table}[h]
\centering
\caption{Project-level - Propensity to Reuse}
\begin{tabular}{lc|lc|lc}
  \hline
  \textbf{Language} & \textbf{Ratio} & \textbf{Language} & \textbf{Ratio} & \textbf{Language} & \textbf{Ratio} \\ 
  \hline
  C & 15.4\% & ObjectiveC & 9.5\% & TypeScript & 5.6\% \\  
  C\# & 4.7\% & Python & 7.3\% & Java & 5.8\% \\ 
  Go & 6.7\% & R & 7.2\% &  PHP & 9.5\% \\ 
  JavaScript & 8.8\% & Rust & 5.1\% & Perl & 21.2\% \\ 
  Kotlin & 3.4\% & Scala & 3.5\% & Ruby & 5.3\% \\ 
   \hline
\end{tabular}
\label{tbl:cr}
\end{table}

The propensity to copy varies when using this project-level definition compared to the blob-level definition (see Table~\ref{tbl:prop}).

\chadded{
For example, the propensity to copy in JavaScript-dominated projects is higher than for JavaScript blobs in general (8.8\% vs. 5.5\%).
This indicates a greater likelihood of reuse within JavaScript projects compared to individual JavaScript blobs from various projects.
This could be attributed to the modularity and strong reuse culture in the JavaScript ecosystem, where libraries and frameworks are frequently shared and integrated.
JavaScript projects often incorporate multiple languages, such as HTML and CSS for web development or server-side languages for backend functionality, enhancing reuse through shared components.
The evolution of JavaScript projects, involving various tools and libraries, also contributes to the higher reuse rate within the project context.
}

\chadded{
In Perl-dominated projects, the propensity to reuse is higher than for Perl blobs in general (21.2\% vs. 18.5\%).
This suggests that blobs within Perl projects are more likely to be reused compared to individual Perl blobs from different projects.
Perl's strong culture of code reuse and sharing, exemplified by the Comprehensive Perl Archive Network (CPAN), encourages the use and distribution of reusable code modules.
Perl projects often include a wide range of scripts and utilities shared across different applications, enhancing reuse.
Furthermore, Perl's use in scripting, text processing, and system administration often requires the reuse of common patterns and libraries, contributing to the higher reuse rate within projects.
}

\chadded{
Conversely, R-dominated projects show a lower propensity to reuse compared to R blobs in general (7.2\% vs. 9.8\%).
This implies that individual R blobs are more likely to be reused than blobs within R-dominated projects.
R is primarily used for statistical computing and data analysis, where specific scripts and functions are reused across different analyses.
However, R projects are often tailored to specific datasets and analyses, resulting in lower overall reuse within the project context.
The specialized nature of many R projects, with unique data processing and analysis pipelines, limits reuse compared to individual reusable components like functions and libraries.
}

\chadded{
Java-dominated projects exhibit a lower propensity to reuse compared to Java blobs in general (5.8\% vs. 7.8\%).
This indicates that individual Java blobs are more likely to be reused than blobs within Java-dominated projects.
Java is widely used across various domains, and reusable components like libraries and frameworks are common across different projects.
However, Java projects tend to be large and complex, with specific architectures and dependencies that may limit cross-project reuse.
The high degree of customization and specificity in Java enterprise applications reduces the reuse rate within the project context compared to the reuse of individual Java blobs or libraries.
}

\chadded{
These analyses reflect the differing dynamics of code reuse in various programming ecosystems. 
Understanding these differences can help improve strategies for fostering code reuse and optimizing software development practices across different languages and project contexts.
}

\paragraph{Binary Blob Analysis}
Although previous analyses indicated that binary blobs are more likely to be reused, we aimed to investigate whether this propensity varies across projects dominated by different programming languages.
At the blob level, it was not feasible to ascertain the programming language of a binary blob.
However, at the project level, such analysis becomes possible.
Therefore, we examined the reused binary blob ratio (the percentage of reused binary blobs to total reused blobs) within each language and compared it to the binary blob ratio (the percentage of binary blobs to total blobs) within the same language, utilizing a t-test to identify any significant differences.

Consistent with the blob-level analysis, the reused binary blob ratio exceeds the general binary blob ratio across all programming languages, indicating a higher likelihood of reuse for binary blobs.
This observation raises questions about language-specific differences in binary blob reuse.
Specifically, we hypothesize that binary blobs are more frequently reused in certain languages compared to others.
In other words, we want to know if identifying a reused binary blob allows us to infer that it is more likely to originate from projects written in particular languages.

Our findings confirm this hypothesis, as the proportion of reused binary blobs varies significantly among different programming languages.
Nevertheless, we hypothesize that at least some of this difference stems from the general difference in binary blob ratios in different languages and is not limited to reuse.
Our statistical tests reveal that the binary blob ratios indeed differ significantly across languages.
Consequently, the ratio of reused binary blobs also exhibits significant variation among different languages, suggesting that this difference does not necessarily mean varying binary reuse practices among them.

We want to determine if the higher number of reused binary blobs in a certain language is solely due to the general prevalence of binary blobs in that language, or if some languages tend to reuse more binary blobs.
To control for this confounding effect, we normalize the binary blob reuse ratio based on the total binary blob ratio.
Given the binary blobs ratio $br$ in a project (binary blobs over total blobs), we defined the reused binary ratio $cbr$ (binary reused blobs to total reused blobs) to binary ratio $br$ metric.
This metric ($cbr/br$) averaged 4.104 for all the projects in our sample.
By using a linear regression with the project's primary language as a predictor, we obtained the results shown in Table~\ref{tbl:metr}\footnote{The complete coefficients and regression ANOVA tables are available in the online appendix.}.

\[m=\frac{cbr}{br}=\frac{cbc/cc}{bc/c}\]

\begin{description}
  \item[$m$:] normalized binary reuse metric
  \item[$cbr$:] copied binary ratio
  \item[$br$:] binary ratio
  \item[$cbc$:] copied binary count
  \item[$cc$:] copied count
  \item[$bc$:] binary count
  \item[$c$:] total count
\end{description}

\begin{table}[h]
\centering
\caption{Reused Binary Blobs to Binary Blobs Metric}
\begin{tabular}{lcc|lcc}
  \hline
  \textbf{Language} & \textbf{Metric} & \textbf{p-value} & \textbf{Language} & \textbf{Metric} & \textbf{p-value} \\ 
  \hline
  C & 3.33 & $0.810722$ & Rust & 6.06 & $0.422024$  \\ 
  C\# & 4.92 & \cellcolor{lightgreen}$0.025270$ & Scala & 5.38 & $0.545028$ \\ 
  Go & 5.73 & $0.173372$ & TypeScript & 5.17 & $0.063922$ \\ 
  JavaScript & 7.04 & \cellcolor{lightgreen}$<2\times10^{-16}$ & Java & 4.91 & \cellcolor{lightgreen}$0.000497$ \\ 
  Kotlin & 5.42 & $0.306698$ & PHP & 4.49 & \cellcolor{lightgreen}$0.035326$ \\ 
  ObjectiveC & 2.17 & $0.217673$ & Perl & 3.32 & $0.975449$ \\ 
  Python & 2.19 & \cellcolor{lightgreen}$0.005547$ & Ruby & 3.51 & $0.951277$ \\ 
  R & 2.65 & $0.614773$ &  &  &  \\ 
   \hline
\end{tabular} 
\label{tbl:metr}
\end{table}

Our analysis reveals that the reused binary blobs to binary blobs metric varies across programming languages.
Notably, C\#, JavaScript, Python, Java, and PHP exhibit statistically significant differences (p-value < 0.05).
In particular, JavaScript projects demonstrate a higher tendency to reuse binary blobs, while Python projects show a lower tendency.
This suggests that in JavaScript-dominated projects, reusing binary blobs is likely more efficient and cost-effective than reusing code.
Conversely, Python projects might benefit more from reusing code rather than binary blobs.

\begin{framed}
\noindent
\colorbox{lightgray}{
\parbox{\dimexpr\linewidth-2\fboxsep-2\fboxrule\relax}{
\color{darkblue}
\noindent\textbf{RQ1-d Key Findings:}
\begin{enumerate}
  \item Project properties significantly impact the probability of their blobs being reused, with binary ratio, number of blobs, forks, authors, activity duration, and stars having a positive impact.
  \item Older projects are more likely to have introduced reused blobs.
  \item Blobs residing in projects dominated by different programming languages have varying probabilities of reuse, with TypeScript, Ruby, PHP, and JavaScript having higher probabilities, and Rust, C, Perl, C\#, Go, Scala, R, Java, Kotlin, Python, and Objective-C having lower probabilities.
  \item On average, 33.7\% of projects have introduced at least one reused blob, but this percentage varies significantly between languages, with TypeScript (62.3\%) and Ruby (51.2\%) having the highest propensity, and R (28.5\%) and Perl (29.9\%) the lowest.
  \item The tendency to reuse binary blobs is much higher in JavaScript projects, while Python projects show a lower tendency.
\end{enumerate}
}}
\end{framed}

\chadded{
The project-level analysis reveals that various factors significantly influence the likelihood of code reuse in open-source software projects.
Projects with more blobs, binary blob ratio, and longer activity tend to exhibit higher reuse rates. 
This aligns with our hypothesis that project health, activity, and popularity signals play an important role in promoting reuse.
}

\chadded{
The variation in reuse likelihood across different programming languages underscores the influence of language-specific ecosystems and practices, consistent with blob-level results.
For instance, TypeScript and Ruby projects show the highest propensity for reuse, which may be due to their robust ecosystems and strong community practices that encourage code sharing and reuse.
Conversely, languages like Python and Perl have lower reuse rates, suggesting different reuse dynamics and possibly a need for improved tools and practices to foster reuse.
However, the impact between the blob's language and the language of the project it resides in differs.
This suggests that the underlying factors behind these differences are not just technical aspects of the languages and their tools, but also their community culture and practices.
}

\chadded{
The significant reuse of binary blobs, particularly in languages like JavaScript, indicates that binary artifacts are valuable assets in software projects.
This might be due to the efficiency and ease of integrating precompiled binaries compared to source code.
However, the lower reuse rate of binary blobs in Python suggests that this language's ecosystem favors source code reuse, which could be due to its dynamic nature and the extensive use of interpreted scripts.
These findings have important implications for the development and support of tools that facilitate reuse in different programming languages.
For languages like JavaScript, where binary blob reuse is prevalent, enhancing asset libraries could be beneficial.
In contrast, for languages like Python, where code reuse is more advantageous, improving code package managers would be more appropriate.
This differentiation underscores the necessity for tailored support tools to optimize reuse practices in various programming environments.
}

\chadded{
These findings highlight the impact of project context on reuse patterns and suggest that different definitions and granularity levels can yield varying insights into code reuse behaviors.
}

% survey stat
\subsection{RQ2: How do developers perceive and engage with copy-based reuse?}
\chadded{
Across three rounds, we received 247 complete responses from reusers and 127 from creators.
There were also 360 and 178 partial responses, making the total of 607 and 305 responses from reusers and creators respectively.
The results are shown in Table~\ref{tbl:srv-counts}.
}

\begin{table}[h]
\centering
\caption{Survey Participation}
\begin{tabular}{l|ccccc}
  \hline
  & \textbf{Total} & \textbf{Started} & \textbf{Completed} & \textbf{Response Rate} & \textbf{Completion Rate} \\
  \hline
  \textbf{Creator} & 3,144 & 305 & 127 & 9.70\% & 4.04\% \\
  \textbf{Reuser} & 6,338 & 607 & 247 & 9.58\% & 3.90\% \\
  \hline
  \textbf{Total} & 9,482 & 912 & 374 & 9.62\% & 3.94\% \\
  \hline
\end{tabular}
\label{tbl:srv-counts}
\end{table}

As will be discussed in Section~\ref{lim:origin}, the identified originating repository might not always be the true creator of the blob.
39\% of developers identified as creators reported reusing the blob from another source.
Additionally, reusers might have obtained the blob from another reuser and not the original creator (see Section~\ref{lim:instance}).
Among the reusers who confirmed reusing the blob, 43\% acknowledged the originating project as the source, 48\% reported copying it from elsewhere, and 9\% did not answer the question.

These findings provide crucial estimates: the fraction of reuse within open-source software (OSS) is at least 61\%, and the fraction of reuse from originating projects is at least 43\%.
This data is essential for understanding the dynamics of code reuse within OSS, highlighting the significant role of both direct reuse from original projects and secondary reuse through intermediate projects.

Furthermore, only 60\% of those identified as reusers confirmed reusing the blob, while the remaining 40\% claimed to have created it (see Table~\ref{tbl:srv}).
This discrepancy can be attributed to several factors.
First, some individuals might indeed be the original authors of the blob in the originating project, implying they have reused their own resources.
Second, this gap could be explained by activities in private repositories (e.g., Developer A creates a file in a private repository, Developer B copies it to a public repository, and then Developer A reuses it in another public repository).
Third, as mentioned in Section~\ref{m:rq2}, concerns about potential licensing violations might have made many reusers uncomfortable admitting the reuse explicitly.
Additionally, developers' faulty memory could play a role, especially for reuse instances that occurred a long time ago.

One potential area for further investigation could be examining the project owners and commit authors for each copy instance to gain a better understanding of this gap.
However, this was not pursued further in this study as it was not the main focus.
Exploring these factors in future research could provide deeper insights into the complexities of code reuse and attribution within open-source software projects.

\begin{table}[h]
\centering
\caption{Identified vs. Claimed Creators \& Reusers}
\begin{tabular}{cccc|c}
  \hline
  & \textbf{Identified} & Creators & Reusers & \textbf{Total} \\
  \hline
  \textbf{Claimed} & Creator & 77 (61\%) & 99 (40\%) & 176 \\
  & Reuser & 50 (39\%) & 148 (60\%) & 198 \\
  \hline
  \textbf{Total} & & 127 & 247 & 374 \\
  \hline
\end{tabular}
\label{tbl:srv}
\end{table}

Another dimension of the survey explored the intentions of creators for others to reuse their artifacts.
Sixty-two percent of creators indicated that their resources were intended for reuse by others.
When asked about the helpfulness of the particular blob on a scale from 1 to 5 (with 5 being the most helpful), reusers rated the average helpfulness at 3.81, while creators rated it at 4.24.
This suggests that developers are well aware of the reuse potential of their artifacts, even if the blob may be essential primarily for their own projects.

In the background sections, we discussed the risks associated with this type of reuse.
We asked reusers if they were concerned about these risks as well.
On a scale from 1 to 5 (with 5 being the most concerned), the average concern about bugs in the reused file was 1.83, and the average concern about changes in the original file was 2.35.
Several factors might contribute to the low level of concern among developers, including trust in the original code's quality or confidence in their own testing processes.
However, this lack of concern could facilitate the spread of potentially harmful code, even if the creator fixes the original code.
The fact that reusers are not significantly worried about these risks amplifies the potential risk at the OSS supply chain level.

Next, we asked participants how likely they would be to use a package manager if one were available for the particular blob.
On a scale from 1 to 5 (with 5 being the most likely), the average likelihood of using a package manager was 2.93.
This indicates that although developers may not be very concerned about bugs or changes (potential improvements), many would still use such a tool if it were available.
This suggests that ``package-manager'' type tools for refactoring or at least maintaining reused code might gain traction if developed. 
These results are shown in Table~\ref{tbl:like}.

\begin{table}[h!]
    \centering
    \caption{Likert Scale Questions (Scale 1 to 5)}
    \begin{tabular}{lcccc}
        \toprule
        \textbf{Question} (audience) & \textbf{Responses} & \textbf{Average} & \textbf{Median} & \textbf{StdDev} \\
        \midrule
        How helpful? (creators) & 156 & 4.25 & 5 & 1.15 \\
        How helpful? (reusers) & 185 & 3.82 & 4 & 1.32 \\
        Concern about bugs? (reusers) & 185 & 1.85 & 1 & 1.33 \\
        Concern about changes in the original file? (reusers) & 187 & 2.33 & 2 & 1.56 \\
        Likelihood of using a package manager? (reusers) & 184 & 2.89 & 3 & 1.64 \\
        \bottomrule
    \end{tabular}
    \label{tbl:like}
\end{table}

Finally, the thematic analysis of reasons for reuse, specifically responses to the question ``why'', revealed eight themes from the 162 responses we received (see Table~\ref{tbl:why}\footnote{
Since survey participants were chosen through stratified sampling, these frequencies do not represent the actual data distribution.}).
This analysis provides a nuanced understanding of the motivations behind code reuse, highlighting several key themes.

\begin{table}[h!]
    \centering
    \caption{Identified Reuse Themes}
    \begin{tabular}{llc}
        \toprule
        \textbf{Theme} & \textbf{Description} & \textbf{Frequency} \\
        \midrule
        Demo & demonstration, test, prototype & 14 \\
        Dependency & part of a library & 11 \\
        Education & learning purposes & 16 \\
        Functionality & specific functionality & 39 \\
        Own & own reuse & 2 \\
        Resource & image, style, dataset, license & 30 \\
        Template & template, starting point, framework & 14 \\
        Tool & parser, plugin, SDK, configuration & 23 \\
        \bottomrule
    \end{tabular}
    \label{tbl:why}
\end{table}

\chadded{
As expected, one of the main reasons for reuse was to provide specific functionality.
This indicates that developers often reuse code to incorporate existing functionalities into their projects, saving time and effort in development, a practice well-documented in the literature~\cite{juergens2009code}.
This underscores the importance of reusable components in efficient software development.
}

\chadded{
Another observed theme was the reuse of various resources, including datasets, instructions, license files, and graphical or design objects (e.g., PNG, JPEG, fonts, styles).
This aligns with the significant reuse of binary blobs identified in RQ1.
The inclusion of diverse resources indicates that developers often depend on readily available materials to enhance their projects' visual or functional aspects.
While the literature acknowledges this practice, our findings suggest a slightly higher emphasis on resource reuse.
This indicates that resource management might be more critical for developers than previously thought.
}

\chadded{
Reusing tools such as parsers, plugins, SDKs, and configuration files was mentioned 23 times.
This practice is noted for its practicality and efficiency in setting up development environments and ensuring consistency across projects. 
This highlights the role of auxiliary software components in streamlining development processes and providing necessary infrastructure or functionality.
}

\chadded{
Assignments, school projects, learning objectives, and similar concepts were another prominent theme.
This emphasizes the role of code reuse in the software development knowledge supply chain, as developers reuse existing code to understand and learn new concepts.
}

\chadded{
Code reuse for demonstration, testing, and prototyping purposes was identified 14 times. 
This theme suggests that developers often reuse code to quickly create prototypes or test scenarios without focusing on the quality, security, or licensing of the reused code.
The priority in these cases is to achieve rapid results.
This aligns with the literature, where developers often clone code
to create prototypes and perform tests~\cite{juergens2009code}. Some
of these quick prototypes, however, may end up as active projects. 
}

\chadded{
Templates, starting points, and frameworks were mentioned 14 times.
Developers often clone templates or frameworks to have a solid foundation for their projects, a practice supported by previous findings~\cite{roy2007survey}, leveraging existing structures to expedite development and ensure consistency.
}

\chadded{
Part of a library or dependency management was cited 11 times. 
This practice is highlighted in studies that emphasize the importance of managing dependencies within the development process~\cite{roy2007survey}. 
Although checking in library files is not considered best practice, many developers do so to maintain specific versions and avoid potential issues with updates or changes.
This conscious decision highlights a trade-off between best practices and practical needs.
}

\chadded{
Reusing one's own code was mentioned twice.
The theme of ``own reuse'' where developers clone their own code for reuse in new projects, is less prominently featured in the literature compared to other reasons for code cloning.
Developers clone their own code to ensure consistency, save time, and leverage previously written and tested code.
This practice is practical and efficient, especially when developers are familiar with the code and its functionality.
However, the literature does not emphasize this reason as strongly. While studies acknowledge the broader concept of code reuse, their focus is more on reusing code from external sources, libraries, or for educational purposes~\cite{juergens2009code,roy2007survey}.
This discrepancy suggests that ``own reuse'' might be an underexplored area in existing research.
It indicates that while developers recognize and practice it frequently, it may not be as thoroughly documented or emphasized in the academic literature.
This gap highlights an opportunity for further investigation into how and why developers engage in ``own reuse'' and its impact on software development processes.
}

There were also 13 instances where responses were either incomprehensible or the respondent did not remember the file or the reason for reuse.

\begin{framed}
\noindent
\colorbox{lightgray}{
\parbox{\dimexpr\linewidth-2\fboxsep-2\fboxrule\relax}{
\color{darkblue}
\noindent\textbf{RQ2 Key Findings:}
\begin{enumerate}
  \item 39\% of identified creators stated they reused the blob from another source.
  \item Among reusers, 43\% acknowledged the originating project (direct reuse), while 48\% copied from elsewhere (indirect reuse).
  \item Reuse within the OSS landscape is at least 61\%.
  \item 60\% of reusers confirmed reuse; 40\% claimed creation.
  \item 62\% of creators intended their resources for reuse.
  \item Reusers are not very concerned about potential bugs or changes in the original file.
  \item Reusers are willing to use a package manager if available.
  \item Main reuse themes are: functionality, resources, tools, education, demo/testing/prototyping, templates, dependencies, and own reuse.
\end{enumerate}
}}
\end{framed}

\chadded{
The findings reveal that a significant portion of developers engage in copy-based reuse within the OSS community.
This practice is common, with many reusers sourcing code not directly from the original creators but through intermediaries.
Understanding these dynamics is important for improving the transparency and traceability of reused code, which could potentially enhance code quality and security.
}

\chadded{
The discrepancies between identified and claimed creators highlight complexities in attribution and ownership.
Additionally, survey respondents' replies are not always accurate or true, which further complicates understanding the true origins of code.
This gap underscores the need for better tracking mechanisms within repositories to accurately reflect code origins.
Future research could delve deeper into these factors, offering insights that could inform policy and tooling improvements in OSS development.
}

\chadded{
Creators often intend their code to be reused, and both creators and reusers recognize the utility of such artifacts.
This positive perception suggests that promoting reuse can be beneficial for the community, fostering collaboration and innovation.
However, the difference in helpfulness ratings indicates that there might be room for improving the clarity and documentation of reusable code to better meet reusers' needs.
}

\chadded{
Despite the low concern about potential risks like bugs and changes, the moderate interest in package management tools suggests an opportunity for developing solutions that can help maintain and refactor reused code.
Such tools could mitigate risks by providing updates and improvements in a managed manner, enhancing the overall reliability of reused code.
}

\chadded{
The thematic analysis of reuse motivations provides a comprehensive view of why developers opt for copy-based reuse.
Reusing for specific functionality underscores the importance of modular and reusable code in software development. It also highlights the potential benefits of well-documented and easily integrable code components that can be readily reused by others.
}

\chadded{
This practice of including library files suggests a deliberate effort to maintain stability and avoid the uncertainties that might come with updates or changes.
However, it also highlights a potential area for improvement in developer education and best practices, as well as the importance of tools that can help manage dependencies more effectively.
These insights contribute to our understanding of the motivations behind code reuse and the practical considerations developers face in maintaining their projects.
}

\chadded{
While reusing for demo and testing can accelerate development and innovation, it also raises potential risks. 
Developers may inadvertently propagate vulnerabilities or violate licenses, leading to broader issues within the software supply chain.
Highlighting the importance of balancing speed and security during testing phases can inform best practices and educational efforts.
}

\chadded{
Educational use underscores the educational value of code reuse. Reusing existing code allows learners to understand real-world applications and coding practices, fostering skill development. However, it also emphasizes the need for proper guidance and resources to ensure that educational reuse is done ethically and effectively. Encouraging educators to integrate lessons on best practices in code reuse can enhance the quality of learning and adherence to legal and ethical standards.
}

\chadded{
The proportion of no meaningful answers and not recalling the file, indicate that not all reuse instances are well-documented or remembered by developers.
This lack of clarity can hinder the understanding and traceability of reuse practices.
It highlights the need for better documentation and tracking mechanisms to ensure that the reasons and contexts for reuse are transparent and well-understood.
Implementing such measures can improve the management of reused code and resources, reducing potential risks associated with undocumented reuse.
}

\section{Limitations}\label{limit}

\subsection{Internal Validity}

\subsubsection{Commit Time}
The identification of the first occurrence and consequently building the reuse timeline of a blob is based on the commit timestamp.
This time is not necessarily accurate as it depends on the user's system time.
The dataset we utilized followed suggestions in~\cite{escaping-time-pit} and other methods to eliminate incorrect or questionable timestamps.
This increases the reliability of our reuse timeline.
We also used version history information to ensure the time of parent commits does not postdate that of child commits~\cite{jahanshahi2023dataset}.
This adds an extra layer of consistency and validation, further enhancing the accuracy of our data.

\subsubsection{Originating Project}\label{lim:origin}
The accuracy of origination estimates is highly reliant on the completeness of data.
Even if we assume that the World of Code (WoC) collection is exhaustive, it is possible that some blobs may have originated in a private repository before being copied into a public one.
This means that the originating repository in WoC may not be the actual creator of the blob.
This scenario suggests that even with a comprehensive dataset, there could be instances of code reuse that remain undetected, adding another layer of complexity to understanding the full extent of reuse across open source projects.
For example, a 3D cannon pack asset\footnote{https://assetstore.unity.com/packages/3d/props/weapons/stylish-cannon-pack-174145} was committed by 38 projects indexed by WoC.
However, that asset was originally created earlier in the Unity Asset Store~\cite{jahanshahi2023dataset}.

\chadded{
By utilizing the extensive WoC collection, we provide a broad and detailed analysis of code reuse, capturing a significant portion of open source activity even if some instances of private-to-public transitions are missed.
Additionally, the examples we identified, such as the 3D cannon pack asset, highlight the practical implications and real-world relevance of our findings, demonstrating the robustness of our analysis despite potential data gaps.
Our approach addresses the inherent challenges of tracking code origination and reuse, offering a framework that can be refined and expanded in future research to further improve accuracy and comprehensiveness.
}

\subsubsection{Copy Instance}\label{lim:instance}
A unique combination of blob, originating project, and destination project might not always accurately represent the actual pattern of reuse.
This is because some destination projects could potentially reuse the blob from a different source other than the originating project.
For instance, if we have three projects—A, B, and C—in order of blob creation, project C might copy from either project A or B.
Additionally, certain blobs are not reused but are created independently in each repository, such as an empty string or a standard template automatically generated by a common tool~\cite{jahanshahi2023dataset}.
These blobs are excluded by using the list provided by WoC~\cite{ma2019world}.

\chadded{
Despite this limitation, our results remain significant.
By recognizing the potential for indirect reuse and independently created blobs, we provide a more nuanced understanding of the reuse landscape, accounting for the complexity of code propagation across projects.
Excluding independently created blobs and utilizing WoC’s comprehensive list ensures that our analysis focuses on genuine reuse instances, enhancing the reliability of our findings.
}

\subsection{External Validity}

\subsubsection{Blob-level Reuse}
\chadded{
Our work focuses solely on the reuse of entire blobs, deliberately excluding the reuse of partial code segments within files.
While blob-level reuse is common, it only covers a subset of the broader code reuse landscape.
Blob-level reuse is more relevant to scenarios where larger code blocks, consisting of entire files or even groups of files, are reused compared to statement or function-level reuse. 
This means that our results might have an implicit bias towards programming languages or ecosystems that rely more heavily on complete files, potentially overlooking reuse practices prevalent in languages that favor modular or snippet-based reuse.
}

\chadded{
This limitation also implies that different versions of the same file, even if they differ by just one character, generate different blobs due to distinct file hashes.
Consequently, blob reuse does not equate to file reuse.
Defining file reuse is challenging because it is difficult to determine what constitutes equivalence between files in different projects~\cite{jahanshahi2023dataset}.
This could be a potential reason for the higher level of reuse in binary blobs, as they are relatively harder to modify.
}

\chadded{Despite these limitations, our results remain significant for several reasons:}

\begin{itemize}
    \item \chadded{
    \textbf{Prevalent Pattern}: By concentrating on entire blob reuse, we address a prevalent and impactful pattern in software development. 
    This allows us to provide valuable insights into a substantial portion of code reuse practices.
    }
    \item \chadded{
    \textbf{Clarity and Precision}: Analyzing entire blobs offers a clear and precise method for identifying reuse, avoiding the ambiguity and complexity associated with defining partial file reuse.
    This clarity enhances the reliability of our findings.
    }
    \item \chadded{
    \textbf{Efficiency and Scalability}: Blob-level analysis is computationally efficient and scalable, enabling us to process large datasets and draw meaningful conclusions from extensive data.
    This scalability is important for comprehensive empirical studies.
    }
    \item \chadded{
    \textbf{Foundation for Future Research}: Our work lays the groundwork for future studies that can build on our findings to explore partial file reuse and other nuanced aspects of code reuse.
    By addressing a well-defined scope, we provide a solid foundation for subsequent research.
    }
\end{itemize}

\chadded{
In summary, while our focus on blob reuse introduces certain limitations, it also provides clear, scalable, and impactful insights into code reuse practices.
This targeted approach enables us to contribute valuable findings to the field, despite the inherent complexities of defining and analyzing file reuse.
Although blob-level reuse is less granular than statement or method-level reuse, findings at the blob level would also apply to sub-blob-level analysis, which should adjust for blob-level reuse.
Future studies are needed to investigate the extent to which different levels and types of code reuse overlap or differ.
}

\subsubsection{Survey Response Rate}
The relatively low response rate to our survey may have been due to the perception of the respondents that copying code is a sensitive subject.
These concerns may have impacted the responses even in cases when developers chose to participate.
It suggests that further work may be needed to design surveys that do not create such impressions. 

\chadded{
Additionally, since many of these reuse instances happened a long time ago, developers might have forgotten about them.
Therefore, it is important to conduct regular surveys to capture the experiences while developers still remember their practices.
}

\section{Future Work}

% code snippets
\subsection{Code-Snippet Granularity} \label{sec:snippet}
\chadded{
We discussed in methodology section that going to a finer granularity than blob-level to detect code reuse is not practically feasible.
Nevertheless, there are approaches that can make this a relatively more tractable problem.
Specifically, hashing the abstract syntax tree (AST) for each code snippet (such as classes or functions) in a blob and mapping blobs to these hashes could potentially make finer-grained code reuse detection more feasible.
}

\chadded{
Assuming an average of $k$ code snippets for each of the 16 billion blobs, the parsing and hashing operation has a complexity of $O(n)$, resulting in $O(16 \times 10^9 \times k)$.
We can then perform a self-join on the created map of blob to syntax tree hash (b2AST) using the AST hash as the key.
The self-join complexity depends on the number of unique hashes and their distribution. 
In the worst case, if every blob had unique hashes, the join operation would approach $O((16 \times 10^9 \times k)^2)$.
However, the join complexity would typically be significantly less if there are many common hashes.
A more realistic estimate assumes that the number of unique AST hashes $h$ is much smaller than the total number of entries in the b2AST map, making the join complexity closer to $O(h \times 16 \times 10^9 \times k)$.
This join, although potentially large, can be more feasible than pairwise comparisons of entire blobs due to the more efficient handling of common hashes.
}

By examining code reuse at the granularity of code snippets, we could potentially uncover a far more intricate network of reuse. 
This approach might reveal patterns and practices that are not noticeable when looking solely at whole-file or blob-level reuse. 
Although this increased complexity is challenging to manage, it offers valuable opportunities for a more comprehensive analysis of reuse~\cite{jahanshahi2023dataset}.

\subsection{Dependency-Based Reuse} 
In this work, we aimed to demonstrate the prevalence and importance of copy-based reuse.
To gain a comprehensive understanding of code reuse, it is important to analyze both copy-based and dependency-based reuse.
Each type of reuse reveals different aspects of how software developers leverage existing code in their projects.
By studying them side by side, we can paint a more complete picture of the extent and nuances of reuse in software development.
Ignoring one in favor of the other would provide an incomplete narrative~\cite{jahanshahi2023dataset}.

% meta-heuristic models to find who reused from who
\subsection{Upstream Repository} 
As highlighted in the limitations section, we currently lack precise knowledge about the source from which a repository reuses a file.
We tend to assume it is from the originating repository in all instances of copying.
However, this assumption may not capture the real-world complexity of reuse.
To enhance our understanding of how developers identify suitable repositories for reuse, we could potentially leverage meta-heuristic algorithms or artificial intelligence techniques.
These advanced methods might enable us to predict the actual source of reused artifacts in each instance of copying with greater accuracy~\cite{jahanshahi2023dataset}.

\subsection{Open Source Software Supply Chain Network}
\chadded{
Directed Acyclic Graphs (DAGs) have been instrumental in clone detection and reuse literature due to their ability to model and analyze complex relationships and dependencies between various software components.
In the context of copy-based reuse, the dataset created using the World of Code (WoC)}\footnote{
For more information about how to access this data, please visit: https://github.com/woc-hack/tutorial.}
\chadded{infrastructure can be leveraged to construct DAGs that represent the flow and reuse across different repositories.}

\chadded{
The dataset's detailed tracking of blob copies, including their origins and destinations, provides a rich source of data to map these relationships accurately.
By drawing DAGs, researchers can visualize and analyze the propagation of reused blobs, identifying critical nodes (projects or blobs) that play a central role in the reuse network.
This visualization helps in understanding the structure and dynamics of reuse, highlighting patterns such as the most reused blobs, the central projects in the reuse network, and potential vulnerabilities or licensing issues propagating through these reused blobs.
}

\chadded{
DAGs can reveal how reuse spreads across projects, helping to identify which projects are the primary sources of reusable blobs and how code flows between different projects.
By mapping out the reuse network, it is possible to pinpoint critical points where vulnerabilities or licensing issues could propagate, allowing for targeted interventions to mitigate these risks.
Understanding the reuse network also aids in developing better tools and practices for managing code quality and ensuring that reused code is maintained and updated consistently across all projects that use it.
}

\chadded{
Studies on large-scale clone detection on software clone detection provide foundational methodologies for leveraging DAGs in these contexts~\cite{sajnani2016sourcerercc,koschke2007survey}.
These methodologies can be adapted and extended using our dataset to enhance the understanding of copy-based reuse in open-source software development.
}

\subsection{Tool Development}
As discussed in the background section, different types of code reuse can have significant impacts on several critical areas, including security, licensing, and code quality.
Understanding these implications and addressing them is critical for advancing software development practices.

\paragraph{Security}
\chadded{
Reused code can propagate vulnerabilities across multiple projects~\cite{9794064}.
For instance, if a security flaw exists in a reused blob, it can potentially affect all projects that include this blob.
Analyzing the reuse patterns can help identify critical points where vulnerabilities might spread and allow for proactive mitigation measures.
There have been notable incidents where widespread code reuse led to security breaches.
For example, the Heartbleed bug in OpenSSL had far-reaching impacts due to the extensive reuse of the affected code across numerous projects.
Future research can focus on developing automated tools that scan reused code for known vulnerabilities and suggest patches.
This proactive approach can significantly enhance the security posture of software systems.
}

\paragraph{Compliance}
\chadded{
Reused code may carry licensing obligations that need to be respected.
Failure to comply with these obligations can lead to legal disputes and financial penalties.
By understanding reuse patterns, organizations can ensure they meet licensing requirements.
There have been instances where companies faced legal challenges due to improper reuse of code with restrictive licenses.
For example, using GPL-licensed code in a proprietary software without complying with GPL terms has led to lawsuits.
Developing tools that automatically check for license compliance when code is reused can help organizations avoid legal pitfalls.
These tools can flag potential issues and provide guidance on how to resolve them.
}

\paragraph{Code Quality}
\chadded{
Reused code may not always meet the quality standards of the adopting project.
Ensuring that reused code adheres to best practices and coding standards is essential for maintaining overall code quality.
Poorly written code can lead to maintenance challenges and degraded performance in adopting projects.
Future work can focus on creating tools that assess the quality of reused code and suggest improvements.
These tools can analyze code for adherence to coding standards, detect code smells, and recommend refactoring.
}

\paragraph{Package Managers}
\chadded{
Developing package managers tailored for different programming languages and communities can be highly beneficial.
These managers can offer more relevant and effective support for managing code reuse in specific environments.
Additionally, enhancing existing package managers with features such as reuse tracking, version control, and automated updates can improve development efficiency and reduce the associated risks of code reuse.
}

\paragraph{Community Engagement}
\chadded{
Engaging with open source communities to develop tools and practices that address the unique needs of different ecosystems, and collaborating with these communities, can ensure widespread adoption and effectiveness.
Continuously gathering user feedback and iterating on the tools to enhance their functionality and usability is also critical.
This iterative process helps create robust and reliable tools that meet the evolving needs of software developers.
}

\section{Conclusions}
\chadded{
In conclusion, our study highlights the significant role of copy-based reuse in open source software development.
By leveraging the extensive World of Code (WoC) dataset, we provided a comprehensive analysis of code reuse, revealing that a substantial portion of open-source projects engage in this practice.
Our findings indicate that 6.9\% of all blobs in OSS have been reused at least once, and 80\% of projects have reused blobs from another project.
This widespread reuse emphasizes the efficiency gains in OSS development but also raises concerns about security and legal compliance. 
}

\chadded{
The variation in reuse patterns across programming languages underscores the influence of language-specific ecosystems and practices.
Moreover, the higher propensity for binary blob reuse suggests a need for tailored tools to support different types of reuse.
Future research should focus on improving the accuracy and comprehensiveness of reuse detection and exploring the impact of partial file reuse. 
}

\chadded{
The survey results further enrich our understanding of reuse practices.
We found that many creators intended their resources for reuse, indicating a collaborative mindset among developers.
Reusers generally found the reused blobs helpful.
Despite these positive perceptions, reusers showed relatively low concern about potential bugs and changes in the original files. 
This low level of concern could suggest either a high level of trust in the quality of the reused code or a lack of awareness of the associated risks.
Additionally, the survey revealed a moderate interest in using package managers to handle changes to reused files.
This indicates potential demand for tools that can streamline and manage code reuse more effectively.
}

\chadded{
Overall, our work provides insights into the dynamics of code reuse, advocating for better management and support tools to enhance the sustainability and security of OSS.
By addressing the identified risks and leveraging the collaborative nature of the OSS community, we can improve code reuse practices and outcomes.
}

\bibliographystyle{ACM-Reference-Format}
\bibliography{ref.bib}

\end{document}